\documentclass[a4paper]{revtex4}
\usepackage[english]{babel}
\usepackage{array,booktabs}% http://ctan.org/pkg/{array,booktabs}
\usepackage{array}    
\usepackage{calc}
\usepackage{pdflscape}
\usepackage{color}
\usepackage{subfigure}
\usepackage{subfloat}
\usepackage[font=small,labelfont=bf]{caption}
\usepackage{graphicx}
\usepackage{amsmath}
\usepackage{setspace}
\usepackage{tabularx}

\usepackage{float}
\begin{document}
\title{Charged Lepton Flavor Violation $\mu \rightarrow e \gamma$ in
$ \mu $-$ \tau $ Symmetric SUSY SO(10) mSUGRA, NUHM, NUGM and NUSM Theories and LHC}

\author{Kalpana Bora, Gayatri Ghosh}
\address{Department of Physics, Gauhati University, Guwahati 781014, India}

%\ead{{\color{blue}kalpana@gauhati.ac.in}}
%\ead{{\color{blue}gayatrighsh@gmail.com}}
%\ead{{\color{blue}vempati@cts.iisc.ernet.in}}

\begin{abstract}
Charged Lepton Flavor Violation (cLFV) processes like $ \mu  \rightarrow e  \gamma $ are rare decay processes that are another signature of physics beyond standard model. These processes have been studied in various models that could explain neutrino oscillations and mixings. In this work, we present bounds on the cLFV decay $ \mu  \rightarrow e  \gamma $ in a $ \mu $-$ \tau $ symmetric SUSY SO(10) theory, using the type I seesaw mechanism. The updated constraints on BR($ \mu  \rightarrow e  \gamma $) from the MEG
experiment, the recently measured value of Higgs mass at LHC, and the value of $\theta_{13}$ from reactor data have been used. We present our results in mSUGRA, NUHM, NUGM, and NUSM models, and the sensitivity to test these theories at the next run of LHC is also discussed.
\end{abstract}
\maketitle

\section{Introduction}
\bigskip
The flavor changing neutral currents (FCNCs) are forbidden in the standard model (SM) of particle physics, at tree level. They are allowed beyond tree level, but highly suppressed by the GIM mechanism \cite{glashow1970}. Flavor mixing in the standard model quark sector is well established, through processes like $K^{0}-\bar{K^{0}}$ oscillations, $ B_{d}-\bar{B_{d}} $ mixing etc. The phenomenon of neutrino oscillations, already proved by experiments, require one to go beyond the standard model. These neutrino oscillations, and hence mixings, are also expected to induce flavor violations in the charged leptonic sector. Theoretically, such cLFV processes could be induced in different theories with BSM particles such as SUSY GUT \cite{Sidori2007}, SUSY seesaw \cite{S.Antusch2006, L.Cabbibi2012, A.Masiero2004}, Little Higgs model \cite{M.Blank2010}, and models with extra dimensions \cite{K.Agashe2006}. In this work we consider cLFV decay $ \mu \rightarrow e  \gamma $, getting contributions from neutrino oscillations and mixings.
\vspace{.2cm}
\par
Many processes involving cLFV decays could be possible such as  $\mu \rightarrow e $, $ \tau \rightarrow \mu  $ or $ \tau  \rightarrow e $ transitions. For $\mu \rightarrow  eee$  decay, an improvement of up to four orders of magnitude is expected \cite{N.Berger2013}, and similarly for $ \mu  \rightarrow  e $ conversions in atomic nuclei improvements are expected \cite{Ed.V,Y.G,R.M.Carey,R.Kutschke,Kurup}. Improvements for $ \mu  \rightarrow e  \gamma $ decay at the next phase of the MEG experiment is
expected to reach BR ( $ \mu  \rightarrow e + \gamma $) $ \leq $   6$\times 10^{-14}$  \cite{Adam, Baldini}. In this work, we have only considered the decay $ \mu  \rightarrow e + \gamma $, asthis is best constrained by experiments. Such experimental
searches and theoretical studies on cLFV can help us constrain the new physics or BSM theories that could be present just above the electroweak scale, or within the reach of the next run of LHC. It is worth mentioning that in the next run of LHC, the center of mass energies are expected to go to 14 TeV \cite{Christoph, candela, F}.
\vspace{.2cm}
\par
SUSY GUTs naturally give rise to tiny neutrino masses via seesaw mechanisms in which significant contributions to cLFVs could come from flavor violations among heavy sleptons. The lepton flavor violation effects could become significant due to radiative corrections to Dirac neutrino Yukawa couplings (DNY), which might arise if the seesaw scale is slightly lower than the GUT scale \cite{L.Cabbibi2012},\cite{Lee, Hle, Borzumati, Hall, Gabb, Hisano, HisanoJ, Rossi, F. Joaqium, FJ, F.Joaq, Arganda, Hirsch, Est, cabb, MH}. Such studies addressing different seesaw mechanisms have been carried out in \cite{L.Cabbibi2012},\cite{Lee, Hle, Borzumati, Hall, Gabb, Hisano, HisanoJ, Rossi, F. Joaqium, FJ, F.Joaq, Arganda, Hirsch, Est, cabb, MH, Hamb}. In \cite{L.Cabbibi2012}, such studies were done in the scenario when neutrino masses and mixings arise due to the type I seesaw mechanism of SUSY SO(10) theory. In this work the Dirac neutrino Yukawa couplings were of the type - $ Y_{\nu} = Y_{u} $ and $ Y_{\nu} = Y_{u}^{diag} U_{PMNS}$, where $ Y_{u} = V_{CKM}Y_{u}^{diag}V_{CKM}^{\dagger}$. Similar studies were done in \cite{Chowd} in the type II seesaw scenario. Charged Lepton Flavor Violation in the SUSY type II seesaw \cite{Schechter, Mohapat, GLazar, Cheng} models have also been studied earlier \cite{ Rossi, F. Joaqium, FJ, F.Joaq, Arganda, Hirsch, Est, cabb, MH}.
\vspace{.2cm}
\par
In this work we carry out studies on cLFV decay ($ \mu  \rightarrow e  \gamma $) using the type I seesaw mechanism in $ \mu $-$ \tau $ symmetric SUSY SO(10) theories \cite{ Joshipura, Fukuyama}, and hence we check the sensitivity to test the observation of sparticles at the next run of LHC \cite{Christoph, candela, F}, in mSUGRA, non-universal Higgs mass (NUHM), non-universal gaugino mass (NUGM) \cite{chak}, and NUSM \cite{S.Bhatt} models. Such studies in NUGM models were done earlier in \cite{Prof}. It may be noted that $ \mu $-$ \tau $ symmetric SUSY SO(10) theory provides a good fit to the observed neutrino oscillations and mixings. In \cite{L.Cabbibi2012}, such studies were
done using the type I seesaw formula, using an older value of  BR($ \mu \rightarrow e \gamma $) \cite{MEG}. We have used the form of Dirac neutrino Yukawa couplings from \cite{Joshipura}, for $ \tan\beta = 10 $, and $M_{GUT}= 2\times 10^{16}$ GeV. The value of the Higgs mass as measured at LHC \cite{Christoph} and global fit values of the reactor mixing angle $ \theta_{13} $ as measured at Daya Bay, Reno \cite{Forero} have been used in this work. In the global data, the octant of the atmospheric angle $ \theta_{23} $ \cite{bora, KB} still needs attention. Some studies on LFV in SO(10) GUTs have also been presented in \cite{Fukuyama, Fuku}.
\vspace{.1cm}
\par
The minimal supergravity model (mSUGRA) is a well motivated model \cite{Chamseddine, R.Barbarei, L.J.Hall, HP} ; for a review, see \cite{R.L, RArnowitt, Ellis, E. Dudas}; for reviews of the minimal supersymmetric standard model, see \cite{X Tata, S. Dawa}
. In mSUGRA, SUSY is broken in the hidden sector and is communicated to the visible sector MSSM fields via gravitational interactions. The generation of gaugino masses
\cite{E.Creme, L.E, pn, RL, L.A} in mSUGRA (N = 1 supergravity) involves two scales $-$ the spontaneous SUGRA breaking scale in the hidden sector through the singlet chiral superfield and the other one is the GUT breaking scale through the non-singlet chiral superfield \cite{Chamseddine, R.Barbarei, L.J.Hall, HP, R.L, RArnowitt, Ellis, E. Dudas} . In principle these two scales can be different. But in a minimalistic viewpoint, they are usually assumed to be the same \cite{Chamseddine, R.Barbarei, L.J.Hall, HP, R.L, RArnowitt, Ellis, E. Dudas}. This leads to a common mass $m_{0}$ for all the scalars, a common mass $M_{1/2}$ for all the gauginos and a common trilinear SUSY breaking term $A_{0}$ at
the GUT scale, $M_{GUT}\simeq2\times10^{16}$ GeV.
\par
Next, we would like to discuss the universal sfermion masses, assumed in the mSUGRA, NUHM, and NUGM models. SO(10) symmetric soft terms essentially mean
boundary conditions close to NUHM. We are working in the framework of SO(10) theories, in which all the matter fields and the right-handed neutrino are present in the same 16-dimensional representation, and, hence, all the matter fields will have the same mass at the high scale. However, the Higgs fields can have a different mass, as they are
not present in the same representation as the matter field. Thus, the boundary terms for the SO(10) theory are consistent with NUHM and mSUGRA (in mSUGRA, all the Higgs fields will be in the same representation). Deviation from NUHM boundary conditions would typically signal a deviation from the SO(10) boundary conditions. Similarly,
it should be noted that NUGM boundary conditions are also derived from SO(10) models. If the hidden sector has representations that are not singlets under SO(10), one can expect
non-trivial gaugino mass boundary conditions. So, to summarize, both NUHM and NUGM are boundary conditions which are a result of assuming SO(10) symmetric boundary
conditions at the GUT scale in two different ways. One can, of course, assume completely non-SO(10) symmetric soft terms at the high scale, but then it would not be compatible
with the present framework. Moreover, as can be seen from results presented 
in Sect. IV, low energy flavor phenomenology is not much affected by these different boundary conditions at the high scales. In this analysis we also carry out cLFV ($\mu\rightarrow e\gamma$) studies in non-universal scalar masses, the NUSM model \cite{S.Bhatt} where the first two generations of scalar masses
and the third generation of sleptons are very massive. Low energy flavor changing neutral current processes (FCNCs) get a contribution due to this non-universality through SUSY
loops. But the requirement of radiative breaking of electroweak symmetry REWSB forbids the scalar masses from being too massive. This circumstance is evaded by allowing
third generation squark masses and the Higgs scalar mass parameters to be small \cite{S.Bhatt}. This smallness also serves to keep the naturalness problem within control. We show the variation of BR($\mu\rightarrow e\gamma$) with $ \frac{m_{1}-m_{2}}{m_{1}+m_{2}}$, where $m_{1}$ and $ m_{2} $ are the masses of the first and second generation sfermions, respectively, in Fig.6. It can be seen that the branching ratio of the cLFV decay $\mu\rightarrow e\gamma$ is not affected much by these completely non-universal SO(10) symmetric mass terms at the GUT scale. It is well known that SUSY can be broken by soft terms of type $- A_{0}, m_{0}, M_{1/2}$ , where $A_{0}$ is the universal trilinear coupling, $m_{0}$ is the universal scalar mass, and $M_{1/2}$ is the universal gaugino mass. Strict universality between Higgs and matter fields of mSUGRA models can be relaxed in NUHM \cite{EllisJ} models. As shown in our results in Sect. IV in mSUGRA, the spectrum of $M_{1/2}$ and $ m_{0} $ is found to lie toward the heavy side, as allowed by MEG constraints on BR($ \mu \rightarrow e \gamma $), though in NUHM, lighter spectra are possible (due to partial cancelations in the flavor violating term). So it motivated us to investigate cLFV decay $ \mu \rightarrow e \gamma $ in NUGM \cite{chak}. Non-universality of gaugino masses can be realized in various scenarios, including grand unification \cite{S.Komine}. In these
models, gaugino masses are non-universal at GUT scales, unlike in mSUGRA/NUHM models. From \cite{chak} we have used
\begin{equation}
M_{1}:M_{2}:M_{3} = -1/2:-3/2:1
\end{equation}
for SO(10) theory. Here, $M_1$, $M_2$ and $M_3$  are the gaugino masses at the GUT scale. In NUGM, an increase in the allowed SUSY soft parameter space is observed, as compared
to mSUGRA and NUHM, which lies within the BR($ \mu \rightarrow e \gamma $) limits of MEG 2013. The BR($ \mu \rightarrow e \gamma $) is found to increase with the increase of $m_{0}$ here, which is opposite to mSUGRA and NUHM. This could be explained due to cancelations between chargino and neutralino contributions \cite{Prof, P.Paradisi}. In the NUGM model, the $|{A_{0}}|$ is found to shift toward the large value side, as compared to mSUGRA and NUHM models. As shown in our results in Sect.IV, we find that in the NUSM model the gaugino masses are very large, so as to allow very large scalar masses. As long as the third generation squark masses and the Higgs scalar mass parameters are small, the fine tuning problem of naturalness does not get worse. In order to have a Higgs mass around 125.9 GeV, the first two generations of squark and slepton masses as well as the third generation of slepton masses lies around 12.5–16 TeV.
\par
From above it is seen that the signatures of cLFV could be tested at the next run of LHC, if the SUSY sparticles are observed within a few TeV, as discussed in more detail
in the next sections. It is worth mentioning here that, during the last run of LHC, no SUSY partner of SM has been observed, and this could point to a high scale SUSY theory. The LHC has stringent limits on the sparticles, which could imply a tuning of EW symmetry at a few percent level \cite{TG, A.Arv, EHard, JL, T. Gher, Fan}. And hence some alternatives to low scale SUSY theories have been proposed. Some of them are minisplit SUSY \cite{A.Aravan} and maximally natural SUSY \cite{Dimopoulos}. In the former the scalar sparticles are heavier than the sfermions (gauginos and higgsinos), so that sfermions could be observed at LHC. Scalar sparticles could be anywhere in the range (10$-10^{5})$ TeV. In maximally natural SUSY, the 4D theories arise from 5D SUSY theory, with Scherk–Schwarz SUSY breaking at a Kaluza–Klein scale $ \sim \frac{1}{R}$ of several TeV \cite{Dimopoulos}. Some aspects of LFV in such theories have been studied in \cite{Garcia}. Charged lepton flavor violation in these models will be studied in our
future work.
\bigskip
\begin{table}[H]
\begin{center}
\begin{tabular}{|c|c|c|}

\hline 
\textbf{LFV Processes} & \textbf{Present Bound} & \textbf{Near Future Sensitivity Of Ongoing Experiments} \\ 
\hline 
%%\br
$ BR \left( \mu  \rightarrow  e \gamma \right)$ & $5.7\times 10^{-13}$ \cite{Adam} &$ 6\times10^{-14}$ \cite{Baldini}  \\ 
 \hline
$BR \left( \tau  \rightarrow  e \gamma \right)$ & $3.3\times 10^{-8}$ \cite{Y. Ahmis}  & $ 10^{-8}-10^{-9}$ \cite{ayasaka} \\ 
 \hline
$BR \left( \tau  \rightarrow  \mu \gamma \right)$ & $4.4\times 10^{-8}$ \cite{Y. Ahmis} &$ 10^{-8}-10^{-9} $\cite{ayasaka}\\
\hline 
$ BR \left( \mu  \rightarrow  eee\right)$ & $ 1.0\times 10^{-12}$\cite{Bellgardt}&$10^{-16} $\cite{Blond}\\
\hline
$BR \left( \tau  \rightarrow  eee \right)$ & $ 2.7\times 10^{-8}$\cite{Hayasaka}&$ 10^{-9}-10^{-10} $\cite{ayasaka}\\
\hline
$BR \left( \tau  \rightarrow  \mu\mu\mu \right)$ &$ 2.1\times10^{-8} $\cite{Hayasaka}&$ 10^{-9}-10^{-10}$\cite{ayasaka}\\
\hline

\end{tabular}
\end{center}
\caption{Present experimental limits and future sensitivities for some LFV processes.}

\end{table}

\bigskip
 
\par 
The paper is organized as follows. In Sect. II, we give connections of cLFV with the type I seesaw mechanism in μ–τ symmetric SO(10) theories. In Sect.{\color{blue} III}, the values of various parameters used in our analysis have been presented. Values of different LFV observables, from \cite{Y. Ahmis, ayasaka, Bellgardt, Blond, Hayasaka}, are listed in Table I. We have used software SuSeFLAV \cite{Garani} to compute BR($ \mu \rightarrow e \gamma $). Section IV contains our results and their analysis. Section V summarizes the work. 
\section{Charged LFV $ \mu \rightarrow e \gamma $ decay in $\mu-\tau$ symmetric SUSY SO(10) theory}
\bigskip
\subsection{cLFV}
\bigskip
Neutrino oscillations and mixings are now a proved phenomenon, and through neutrino oscillations, a cLFV process could be induced as
\begin{equation}
l_{i}\xrightarrow{W}\nu_{l_{i}}\rightarrow\nu_{l_{j}}\xrightarrow{W}l_{j}
\end{equation}
Here W means a vertex involving a W boson. The process requires neutrino mass insertion at two points. In the type I seesaw mechanism, $\Delta$L = 2 Majorana neutrino masses arise from tree level exchange of a heavy right-handed neutrino. The SUSY SO(10) theory naturally incorporates the seesaw mechanism. The presence of heavy RH neutrinos at an intermediate scale leads to the running and generates flavor violating entries in the left-handed slepton mass matrix at the weak scale \cite{L.Cabbibi2012}. The lepton flavor violating entries in the SO(10) SUSY GUT framework can be understood in terms of the low energy parameters. These entries in the leading log approxi-
mation in mSUGRA are \cite{Cabbibi}. 
\begin{equation}
 \left( m^{2}_{\tilde{L}} \right)_{i\neq j} = \frac{-3m_{o}^{2}+A_{o}^{2}}{ 8\pi^{2}} \sum_{k}
      \left(Y_{\nu}^{\star}\right)_{ik}\left(Y_{\nu}\right)_{jk} \\ log\left(\frac{M_{X}}{M_{R_{k}}}\right) 
\end{equation}
Here $M_{X}$ is the GUT scale,  $M_{R_{k}}$ is the scale of the $k^{th}$ heavy RH majorana neutrino, $m_{0}$ and $A_{0}$ are universal soft mass and trilinear terms at the high scale.  $ Y_{\nu} $ are the Dirac neutrino Yukawa couplings. The flavor violation is parameterized in terms of the quantity $\delta_{ij}=\frac{\Delta_{ij}}{\overline{m}^{2}_{\tilde{l}}}$ . Here ${{\overline{m}}}^{2}_{\tilde{l}} $ is the geometric mean of the slepton squared masses \cite{Gabb}, and $ \Delta_{i \neq j}$  are flavor non-diagonal entries of the slepton mass matrix induced
at the weak scale due to RG evolution. The mass insertions are branched into the LL/LR/RL/RR types \cite{L.Sa}, according to the chirality of the corresponding SM fermions. The fermion masses can be generated by renormalizable Yukawa
couplings of the 10$\oplus$126$\oplus\overline{120}$ representation of scalars of
SO(10) GUTs. We have used the Dirac neutrino Yukawa couplings $ Y_{\nu} $ at the high scale in $\mu$-$\tau$ symmetric SO(10) GUTs in our work from \cite{Joshipura},
\begin{equation}
 Y_{\nu}=\frac{1}{\upsilon sin\beta}M_{D}
\end{equation}
$M_{D}$ is the Dirac neutrino mass matrix. The flavor violating off-diagonal entries at the weak scale in Eq. (3) are then completely determined by using $Y_{\nu}$ from Eq. (4). To calculate the $ \delta $s from the RGEs, we use the leading log approximation. Assuming the soft masses to be flavor universal at the input scale, off-diagonal entries in the LL sector are induced by right-handed neutrinos running in the loops. To use the leading log expression (Eq.3) we need the mass of the heaviest right-handed neutrino, which we have used from \cite{Joshipura} by diagonalizing the matrix $ M_{R} $ , and it is found to be $\sim 10^{16} $ GeV. The induced off-diagonal entries relevant to  $l_{i}$ $\rightarrow$ $l_{j}$ +$ \gamma $ are of the order of (putting $ A_{0} $ to 0),
\begin{equation}
\left( \delta_{LL}\right) _{\mu e} = \frac{-3}{8\pi^{2}} \left( Y_{\nu}^{\star}\right) _{13}\left( Y_{\nu}\right) _{23}ln\left( \frac{M_{X}}{M_{R_{3}}}\right) 
\end{equation}

\begin{equation}
\left( \delta_{LL}\right) _{\tau \mu} = \frac{-3}{8\pi^{2}} \left( Y_{\nu}^{\star}\right) _{23}\left( Y_{\nu}\right) _{33}ln\left( \frac{M_{X}}{M_{R_{3}}}\right) 
\end{equation}

\begin{equation}
\left( \delta_{LL}\right) _{\tau e} = \frac{-3}{8\pi^{2}} \left( Y_{\nu}^{\star}\right) _{13}\left( Y_{\nu}\right) _{33}ln\left( \frac{M_{X}}{M_{R_{3}}}\right) 
\end{equation}

\begin{table}[H]

\begin{center}

\begin{tabular}{|c|c|}

\hline 
\textbf{LFV contributions} &\textbf{For $\mu$-$\tau$ symmetric case} \\ 
\hline 
%%\br
$\delta_{12}$ & $0.9519\times 10^{-3}$  \\ 
 \hline
$\delta_{23}$ & $3.488\times 10^{-4}$ \\ 
 \hline
$\delta_{31}$ & $0.92308\times 10^{-3}$ \\ 
\hline

\end{tabular}
\end{center}
\caption{Values (dominant) of $\delta_{ij}$ that enter Eq. (5,6,7)
for $\mu$-$\tau$ symmetric theory.}
\end{table}

\bigskip
The branching ratio of a charged LFV decay $ l_{i}$ $\rightarrow $ $l_{j}$ is \cite{L.Cabbibi2012},
\begin{equation}
\text{BR} \left( l_{i} \rightarrow l_{j}+\gamma \right)\approx \alpha^{3}\frac{\vert\delta^{LL}_{ij}\vert^{2}}{G_{F}^{2}M^{4}_{SUSY}}\tan^{2}\hspace{.01cm}\beta \linebreak \text{BR} \left( l_{i}  \rightarrow  l_{j}\nu_{i}\tilde{\nu_{j}} \right)
\end{equation}
where $ M_{SUSY } $ is the SUSY breaking scale. In NUHM models, the term  $(-3m_{o}^2+A_{o}^2)$ of the mSUGRA models in Eq. (3) is replaced by $(-2m_{o}^2+A_{o}^2+m^2_{H_u})$. Here, $ m_{H_u} $ is the  soft mass terms of the up type Higgs at the high scale. We consider the NUHM1 case (at the GUT scale) 
\begin{equation}
m_{H_{u}}=m_{H_{d}}
\end{equation}
Moreover, there can be a relative sign difference between the universal mass terms for the matter fields and the Higgs mass terms at the GUT scale. This can clearly lead to cancellations
for
\begin{equation}
m^{2}_{H_{u}} \approx -2m_{0}^{2}
\end{equation}
Or enhancements for
\begin{equation}
m^{2}_{H_{u}} \geq m_{0}^{2}
\end{equation}
compared to mSUGRA in the flavor violating entries at the weak scale.
\bigskip
\subsection{\textbf{$\mu-\tau$ symmetry}}
\bigskip
Neutrino mixings observed in various oscillation experiments can be explained through the structure of both the neutrino and the charged lepton mass matrices. In a basis
where the charged leptons are mass eigenstates, the $\mu \leftrightarrow \tau$
interchange symmetry has proved useful in understanding the experimentally observed near-maximal value of $\nu_{\mu}\leftrightarrow\nu_{\tau}$  mixing angle ($\theta_{23}\simeq \frac{\pi}{4}$). In $\mu-\tau$ symmetry the mass matrix remains invariant under the interchange of the 2–3 sector. The mass matrix becomes
\begin{equation}
M_{\nu} = \begin{pmatrix}
x & a & a\\
a & b & c\\
a & c & b\\
\end{pmatrix}
\end{equation}
At high scales the $\mu \leftrightarrow \tau$ symmetry can be assumed to be an exact symmetry. But at low scales the $\mu \leftrightarrow \tau$ symmetry is effective only in the neutrino Yukawa couplings but not in the charged lepton sectors, since $m_{\tau}>m_{\mu}$ . An immediate consequence of this class of theories is that $\theta_{23} = \frac{\pi}{4}$ and $\theta_{13} = 0$. Recently evidence of $\theta_{13} \neq 0$  from reactor experiments \cite{DayaBay, Reno, Double} has been found. This reduces $\mu-\tau$  symmetry to an approximate symmetry. A small, explicit, tiny breaking of the $\mu-\tau$  symmetry, to explain the reactor angle $\theta_{13}$, has been studied in \cite{Joshipura} . This can be done by adding a 120-dimensional Higgs to the 10+$\overline{126}$ representation of Higgs. Yukawa interactions of the model are given by the lagrangian
\begin{equation}
-\textit{L}_{\textit{Y}} = 16_{i}[H_{ij}10 + F_{ij}\overline{126} + G_{ij}120]16_{j} + h.c
 \end{equation}
with $H_{ij} = H_{ji}; F_{ij} = F_{ji}; G_{ij} = -G_{ji}$. It may be noted that some results on neutrino masses and mixings using updated values of running quark and lepton masses in SUSY SO(10) have also been presented in \cite{GGhosh}. Though
we consider 3-flavor neutrino scenario, 4-flavor neutrinos with sterile neutrinos as fourth flavor are also possible \cite{PG}. We have not considered the CP violation phase \cite{ddutta}, in this work.
\section{Calculation of BR($\mu \rightarrow e\gamma$) in mSUGRA, NUHM, NUGM, and NUSM}
\bigskip
In this section we present our calculations and results on the charged LFV constraints in  $ \mu $-$ \tau $ symmetric SO(10) SUSY theory, with the type I seesaw mechanism using the NUHM, mSUGRA, NUGM, and NUSM like boundary conditions through detailed numerical analysis. We scan the soft parameter space for mSUGRA in the following ranges:
$$ tan\beta=10 $$
$$ m_{h} \in \left[ 122.5, 129.5\right] \hspace{.1cm}\text{GeV}$$
$$ \Delta m_{H} \in 0 $$
$$ m_{0} \in \left[ 0, 7\right] \hspace{.1cm}\text{TeV} $$
$$ M_{1/2} \in \left[ 0.3, 3.5\right] \hspace{.1cm} \text{TeV}$$
$$ A_{0} \in \left[ -3m_{0} , +3m_{0} \right]$$
\begin{equation}
sgn\left( \mu\right) \in\lbrace-,+\rbrace 
\end{equation}
\vspace{.05cm}
We perform random scans for the following range of parameters in NUGM model with non-universal and opposite sign gaugino masses at $M_{GUT}$, with the sfermion masses
assumed to be universal maintaining the ratio between the non-universal gaugino masses \cite{chak}
$$ m_{h} \in \left[ 122.5, 129.5\right]\hspace{.1cm} \text{GeV} $$
$$m_{0} \in \left[ 0, 7\right]\hspace{.1cm}\text{TeV}$$
$$M_{1} \in \left[ -.3, -2.8\right]\hspace{.1cm}\text{TeV}$$
$$M_{2} \in \left[ -.9, -8.4\right]\hspace{.1cm} \text{TeV}$$
$$M_{3} \in \left[ .6, 5.6\right]\hspace{.1cm} \text{TeV}$$
$$ tan\beta=10 $$
\begin{equation}
A_{0} \in \left[ -3m_{0} , +3m_{0} \right] 
\end{equation}
Here $ m_{0} $ is the universal soft SUSY breaking mass parameter for sfermions, and $ M_{1} $, $ M_{2} $, and $ M_{3} $ denote the gaugino masses for $U(1)_{Y}$, $SU(2)_{L}$ and $SU(3)_{C}$  respectively. $ A_{0} $ is the trilinear scalar interaction coupling, $ \text{tan}\beta $  is the ratio of the MSSM Higgs vacuum expectation values (VEVs).
\par 
We have done the numerical analysis using the publicly available package SuSeFLAV \cite{Garani}. We also study cLFV for the non-universal Higgs model without completely universal soft masses at a high scale. The ranges of the scan of various
SUSY parameters, used by us, in NUHM are:
$$ m_{h} \in \left[ 122.5, 129.5\right]\hspace{.1cm} \text{GeV} $$
$$ 30\hspace{.1cm} \text{GeV} \leq m_{0} \leq 6\hspace{.1cm} \text{TeV} $$
$$30\hspace{.1cm} \text{GeV} \leq M_{1/2} \leq 2.5\hspace{.1cm}\text{TeV}$$
$$- 8.5 \hspace{.1cm}\text{TeV} \leq m_{H_{u}} \leq +8.5 \hspace{.1cm}\text{TeV}$$
$$-8.5\hspace{.1cm}\text{TeV} \leq m_{H_{d}} \leq +8.5 \hspace{.1cm}\text{TeV} $$
\begin{equation}
-18 \hspace{.1cm}\text{TeV} \leq A_{0} \leq + 18\hspace{.1cm}\text{TeV}
 \end{equation}
\par 
The $\Delta^{LL}_{i\neq j}$ due to non-universal Higgs and $ m_{h} $ $\geq$ 125 GeV
puts a strong constraint on SUSY parameter space. Also, because of partial cancelations in the entries of $\Delta^{LL}_{i\neq j}$ in the NUHM case, a large region of parameter space can be explored by MEG. We also perform random scans for the
following range of parameters in NUSM model \cite{S.Bhatt} and generate the SUSY particle spectrum. The ranges of the SUSY parameters used at GUT scale are:
$$ tan\beta=10 $$
$$ m_{0} \in \left[ 0, 16\right] \hspace{.1cm}\text{TeV} $$
$$ M_{1/2} \in \left[ 0, 6\right] \hspace{.1cm}\text{TeV} $$
$$ A_{0}  = 0\hspace{.1cm} \text{TeV} $$
\begin{equation}
 m_{H_{u}}=m_{H_{d}} = 0\hspace{.1cm} \text{TeV} 
\end{equation}
The masses of the heavy neutrinos used in our calculations are - $ M_{R_{1}}=10^{13} \hspace{.1cm}\text{GeV}$, $M_{R_{2}} = 10^{14}\hspace{.1cm}\text{GeV}$, and $M_{R_{3}}=  10^{16}\hspace{.1cm}\text{GeV}$.. For $ \Delta m^{2}_{sol} $, $\Delta m^{2}_{atm}$ and $\theta_{13}$ , we use the central values from the recent global fit of neutrino data \cite{Forero}. The present limits on different LFV observables are summarized in Table I. In Table II we have given the dominant values of $\delta_{ij}$ that enter Eqs. (5), (6), and (7).
\section{Analysis and discussion of results}
\bigskip
In this section, we will present analysis and discussion of results obtained in Sect. III.
\subsection{Complete universality: cMSSM (mSUGRA)}
\bigskip
In mSUGRA at the high scale, the parameters of the model are $m_{0}$, $A_{0}$, and the unified gaugino mass $ M_{1/2} $. In addition to these, there is the Higgs potential parameter $ \mu $ and the undetermined ratio of the Higgs VEVs, tan$\beta$. The entire supersymmetric mass spectrum is determined once these parameters are given. We find that the updated MEG limit \cite{Chowd} together with a large $\theta_{13}$ \cite{Forero} puts significant constraints on the SUSY parameter space in mSUGRA. As can be seen from Fig. 1a, only a small part of the paramater space survives for tan $ \beta $ = 10 in mSUGRA allowed by the future MEG limit for BR($ \mu \rightarrow e \gamma $). This leads to the conclusion that the parameter space $M_{1/2} \geq $ 1 TeV is allowed by present MEG bounds on BR($ \mu \rightarrow e \gamma $), while the future MEG limit excludes a small $M_{1/2}$ space $ \leq$ 3.5 TeV. The allowed regions in Fig. 1b require very heavy spectra, i.e. $ m_{0} $ $ \geq $ 6 TeV for small $M_{1/2}$ or $M_{1/2}$ $ \geq $ 2 TeV for small $ m_{0} $. In Fig. 1c, d we plot the lightest Higgs mass, $ m_{h} $, as a function of $ m_{0} $, $M_{1/2}$ in the mSUGRA case. We see that for the range of the Higgs mass as given by the data at LHC, i.e. i.e 122.5 $\text{GeV} \le m_{h} \le$ 129.5 $\text{GeV}$, $m_{0} \geq$ 1 TeV is allowed by the present MEG bounds on BR($ \mu \rightarrow e \gamma $). The space $M_{1/2} \geq$ 1 TeV is allowed as can be seen from Fig. 1d. In Fig. 1e we have presented the results for the decay $\mu \rightarrow eee $. In SUSY (with conserved R-parity) the dominant contribution to this process arises from the same dipole operator responsible for $\mu \rightarrow e \gamma $ (for R-parity in SUSY theories, see Refs. \cite{Farrar, Dimopo, Weinberg, Sakai, Gell-Mann, T.yanagida, Mohapatra, Babu}). Such a prediction is consistent with our results shown in Fig. 1e for $tan\beta$ = 10. In SUSY with conserved R-parity the two processes $\mu \rightarrow e \gamma $ and $\mu \rightarrow eee $ are correlated. This correlation is clearly seen in Fig. 1e as as $\text{BR} (\mu \rightarrow e\, \gamma) \sim \alpha_{em} \text{BR} (\mu \rightarrow 3\,e)$. Here $ \alpha_{em} $ is the electromagnetic dipole operator. The asymmetry in the value of $A_{0}$ can be seen in Fig. 1f.

\begin{figure}
\begin{subfigure}{\includegraphics[height=6cm,width=7.95cm]{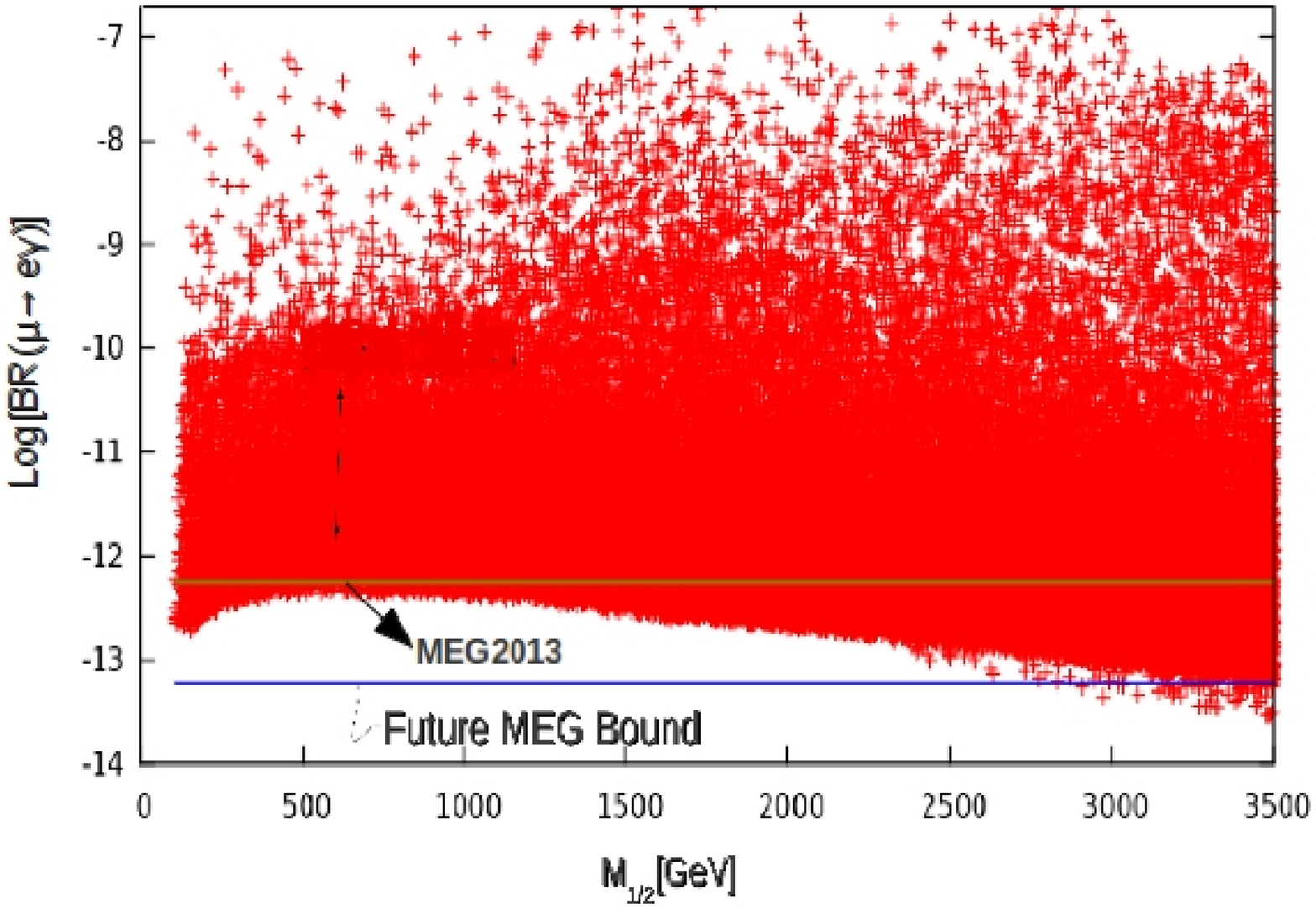}}\end{subfigure} 
\begin{subfigure}{\includegraphics[height=6cm,width=7.95cm]{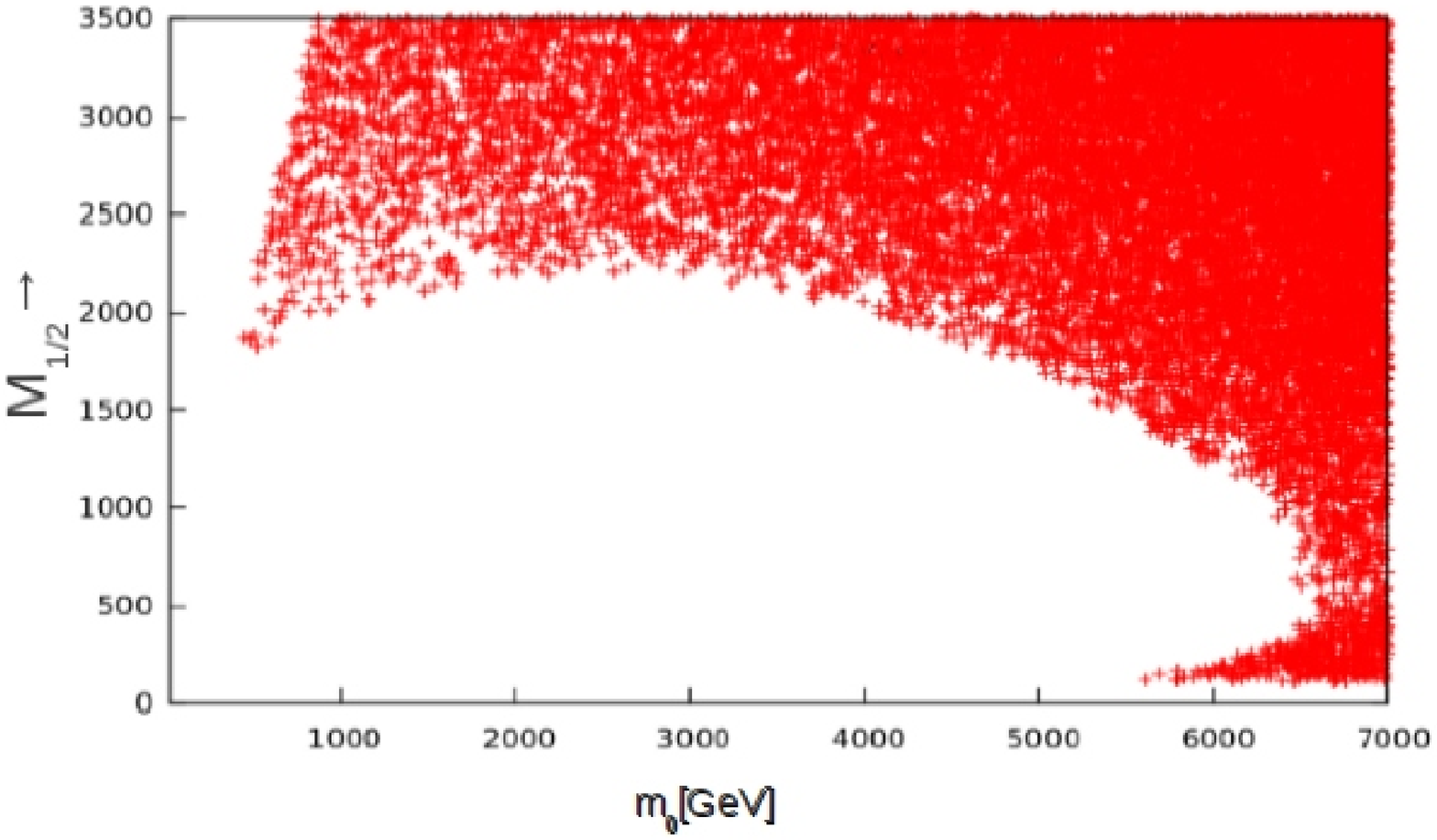}}\end{subfigure}\\
\begin{subfigure}{\includegraphics[height=6cm,width=7.95cm]{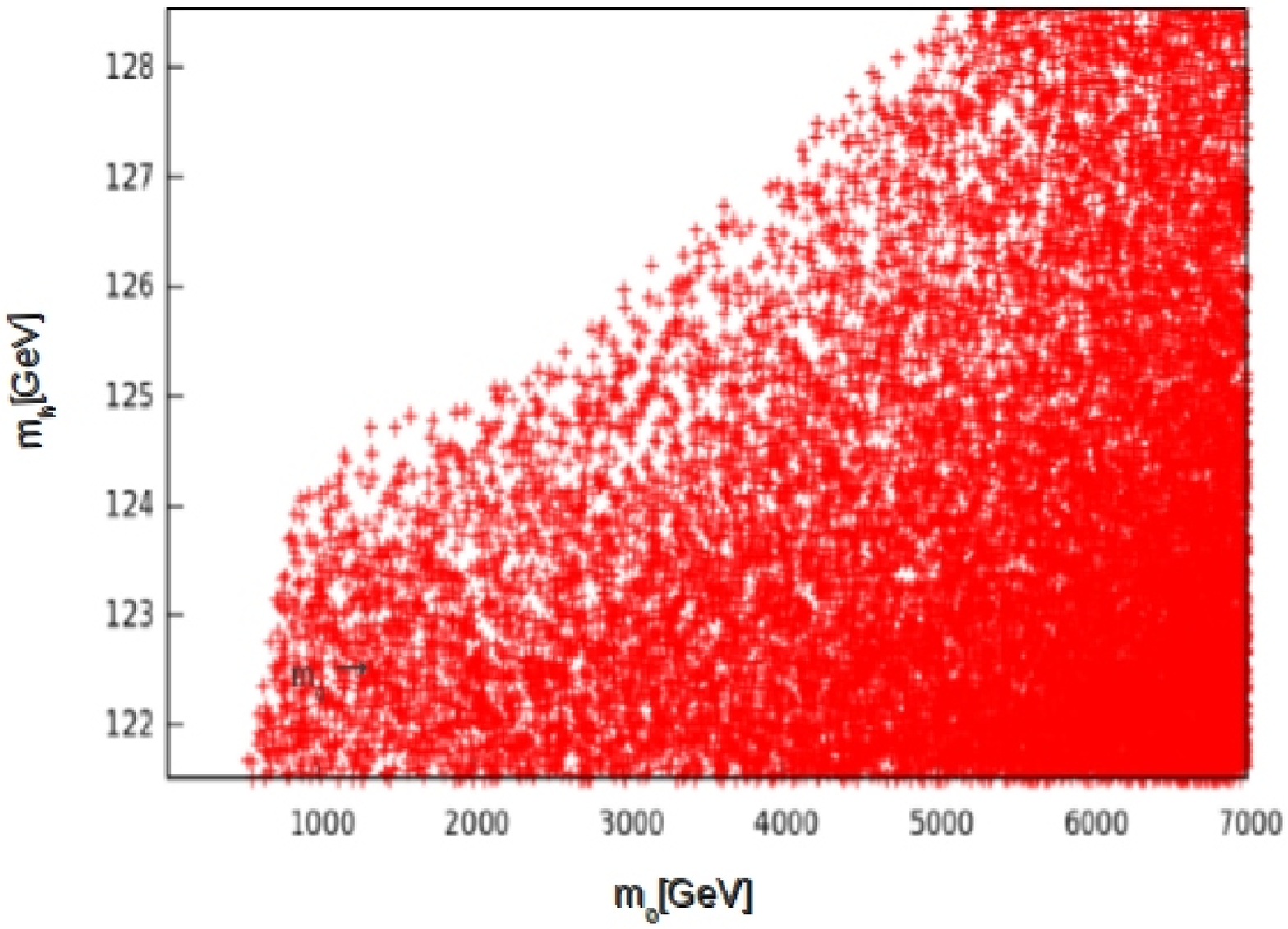}}\end{subfigure} 
\begin{subfigure}{\includegraphics[height=6cm,width=7.95cm]{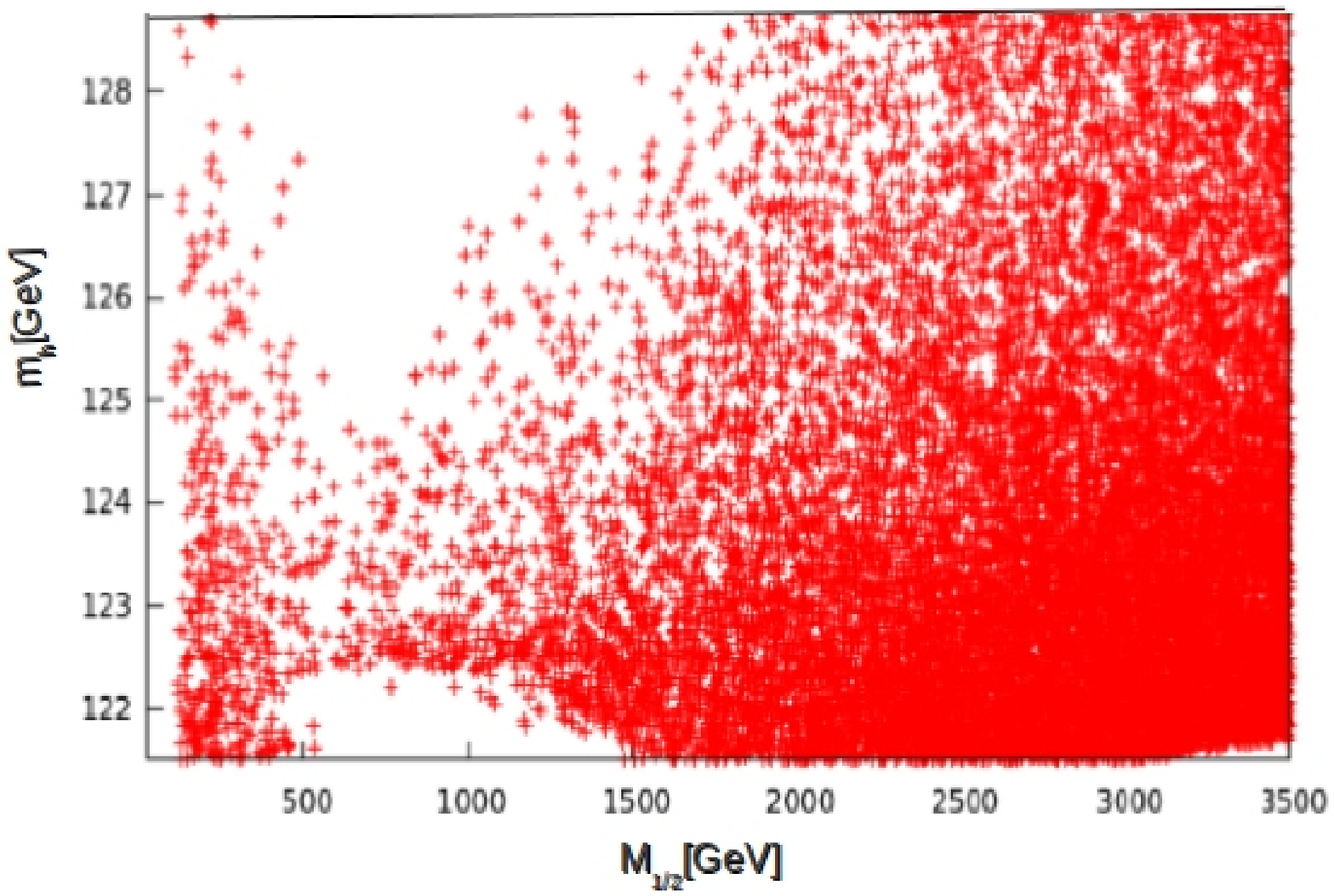}}\end{subfigure}\\
\begin{subfigure}{\includegraphics[height=6cm,width=7.95cm]{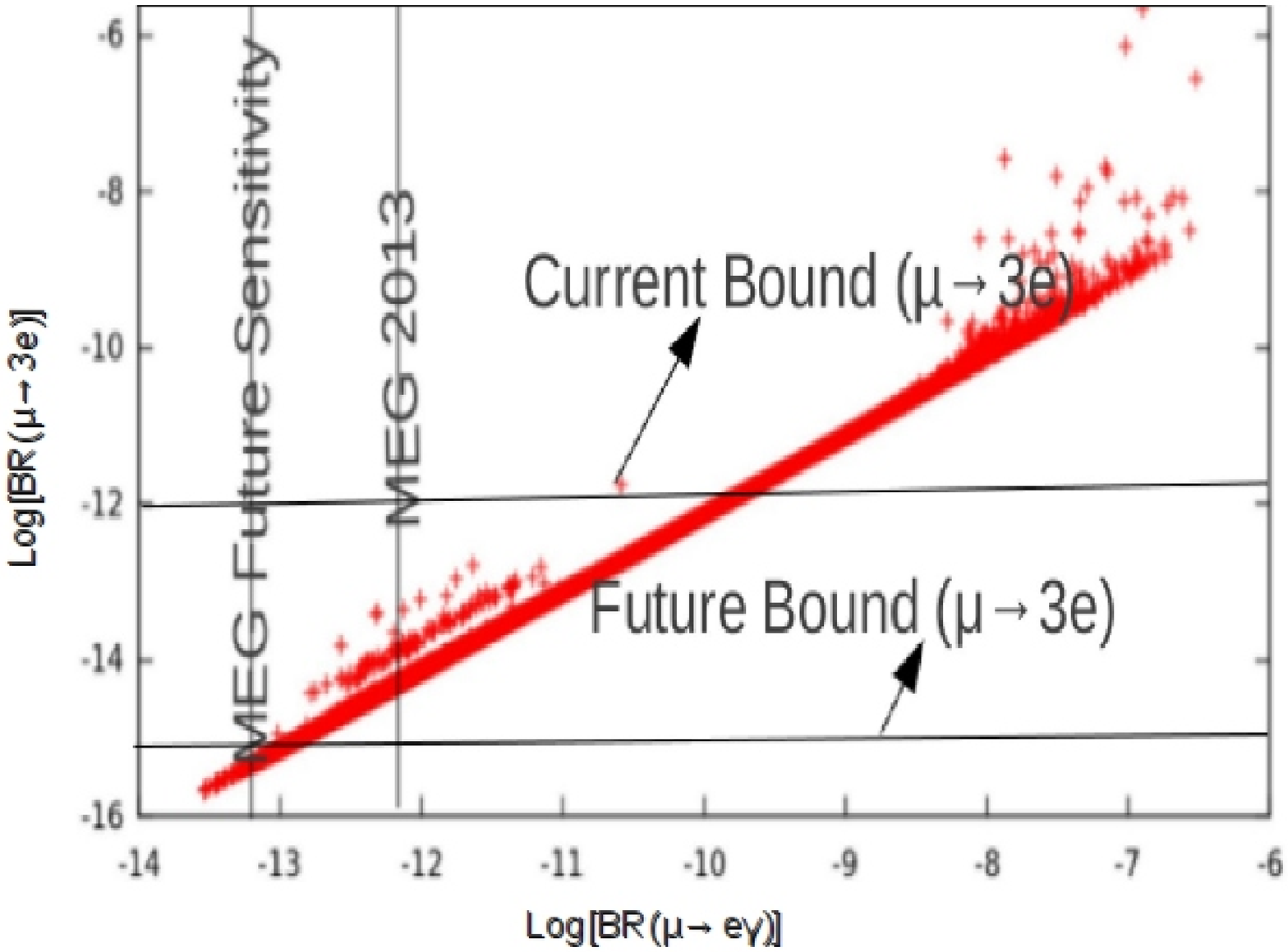}}\end{subfigure} 
\begin{subfigure}{\includegraphics[height=6cm,width=7.95cm]{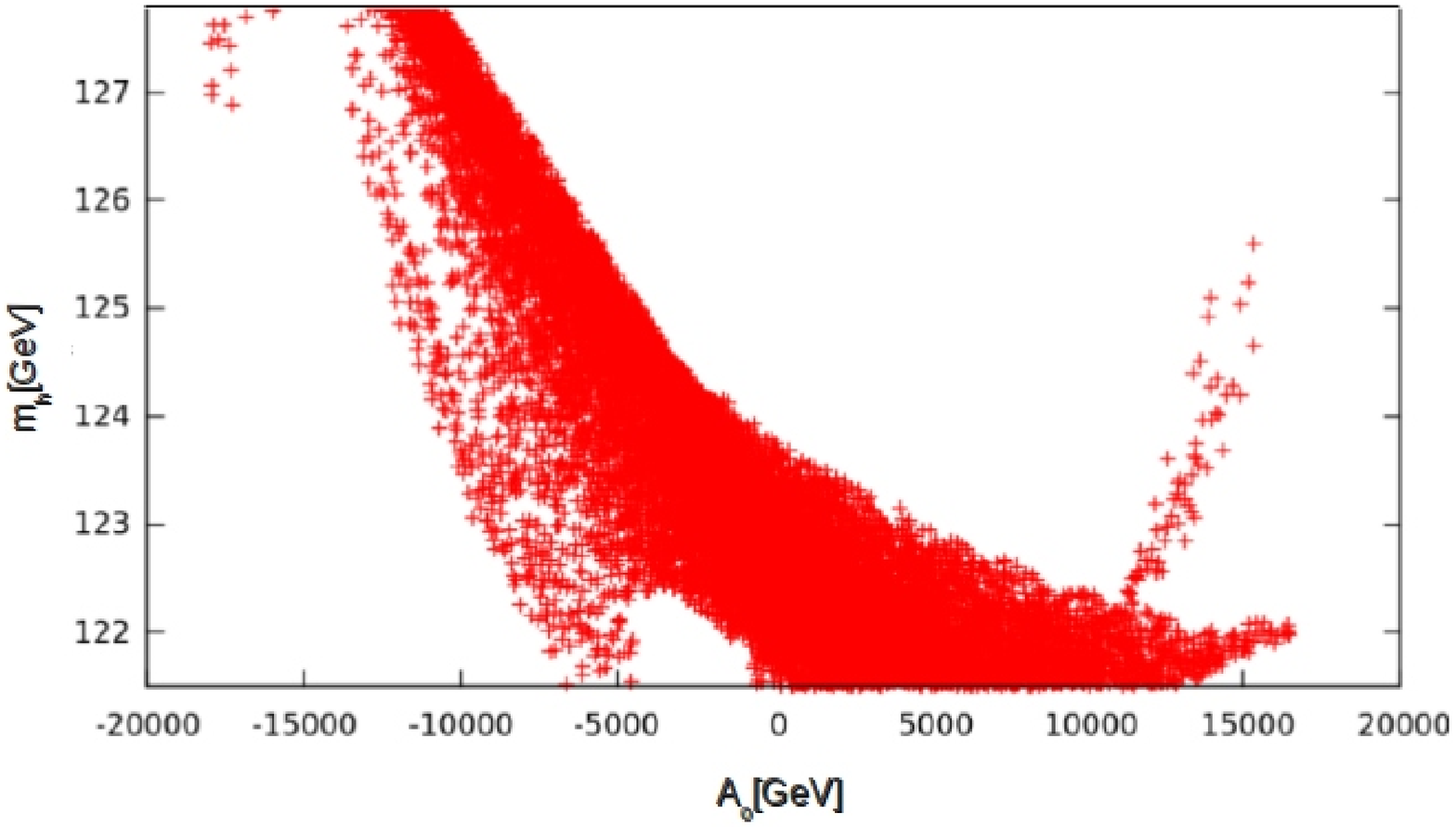}}\end{subfigure}\\
\caption{The results of our calculations are presented for mSUGRA case. In \textbf{a}, different horizontal lines represent the present (MEG 2013) and future MEG bounds for BR($ \mu $ $ \rightarrow $ e + $ \gamma $). \textbf{b-d} SUSY parameter space allowed by MEG 2013 bound.}
%\label{some example}
\end{figure}

\subsection{Non Universal Higgs Model (NUHM1)}
\bigskip
Next, we present our results obtained in NUHM1 case. In Fig. 2a we have shown $M_{1/2}$ vs. log[BR($ \mu \rightarrow e+\gamma$) and the Fig. 2b in the right panel shows $m_{0}$ [GeV] vs. $ M_{1/2} $[GeV]. Different horizontal lines in Fig. 2a correspond to present and future bounds on BR($ \mu $ $ \rightarrow $ e + $ \gamma $). We can see from Fig. 2a, b that even in the presence of partial cancelations, most of the NUHM1 parameter space is going to be explored by present and future bounds of MEG.
\par 
In Fig. 2c, d, the SUSY parameter space $M_{1/2} - m_{h} $ and $m_{0} - m_{h} $ is presented, as allowed by present MEG bounds. For a Higgs mass around 126 GeV, almost all values of $ M_{1/2} $ are allowed in the range (100–2500 GeV). Similarly
for $ m_{h} $ around 126 GeV, the region 3 TeV $ \leq m_{0}\leq $ 6 TeV is
mostly favored. In $ \delta^{LL}_{i \neq j} $ due to cancelations between $ m^{2}_{H_{u}} $ and $ m_{0}^{2} $, a large region of soft parameter space is allowed which
would be easily accessible at the next run of LHC satisfying the current cLFV constraints. Figure 2e shows  $A_{0}$ vs. $ m_{h} $ [GeV]. $A_{0}$ is slightly more symmetric compared to mSUGRA.
\bigskip

\subsection{NON UNIVERSAL GAUGINO MASS MODELS (NUGM)}
\bigskip
From the studies in mSUGRA and NUHM model in the above subsections, we see that the SUSY parameter space, as allowed by future MEG bounds on BR($\mu \rightarrow e+ \gamma$)
shifts to the heavier side. Hence, we are motivated to do such studies in NUGM models. In this section we discuss the scenario with non-universal and opposite sign gaugino masses at $M_{GUT}$ , with the sfermion masses assumed to be universal. We perform random scans for ranges of the parameters given in Eq. (15). We concentrate on the specific model 24 of \cite{chak}  with the gaugino masses having the ratios  $ M_{1}:M_{2}:M_{3}=-1/2:-3/2:1 $. In fact, the non universality of the gaugino masses is by no means a peculiar phenomenon, rather it is realized in various scenarios, including some approaches to grand unification \cite{S.Komine}. Figure 3a reveals that the light spectrum accessible at LHC can be explored by current and future bounds of the MEG Limit. The resulting preference for light sparticle masses renders the detection of NUGM at the LHC operating at 14 TeV collision energy positive.
\begin{figure}
\begin{subfigure}{\includegraphics[height=6cm,width=7.9cm]{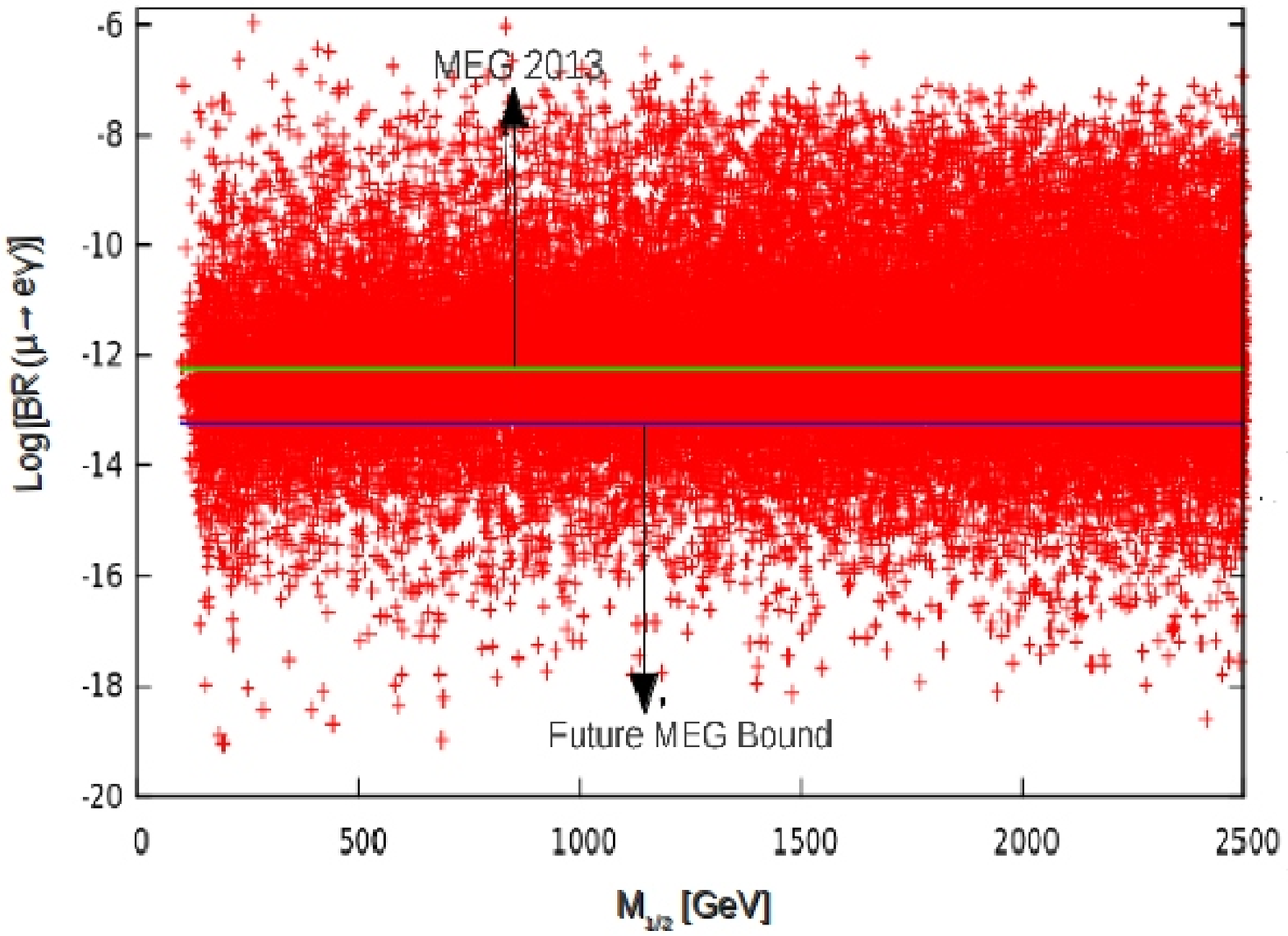}}\end{subfigure} 
\begin{subfigure}{\includegraphics[height=6cm,width=7.9cm]{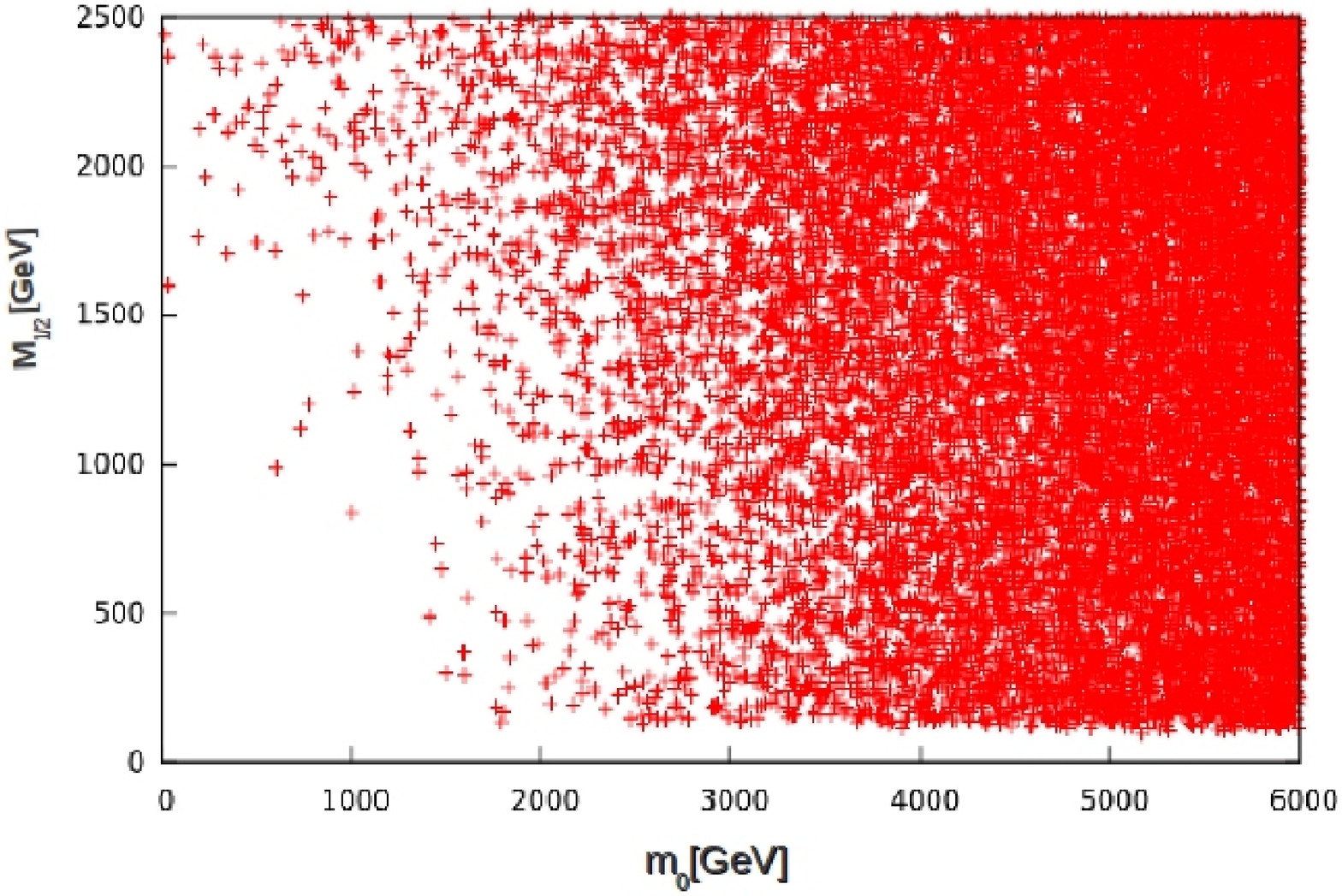}}\end{subfigure}\\
\begin{subfigure}{\includegraphics[height=6cm,width=7.9cm]{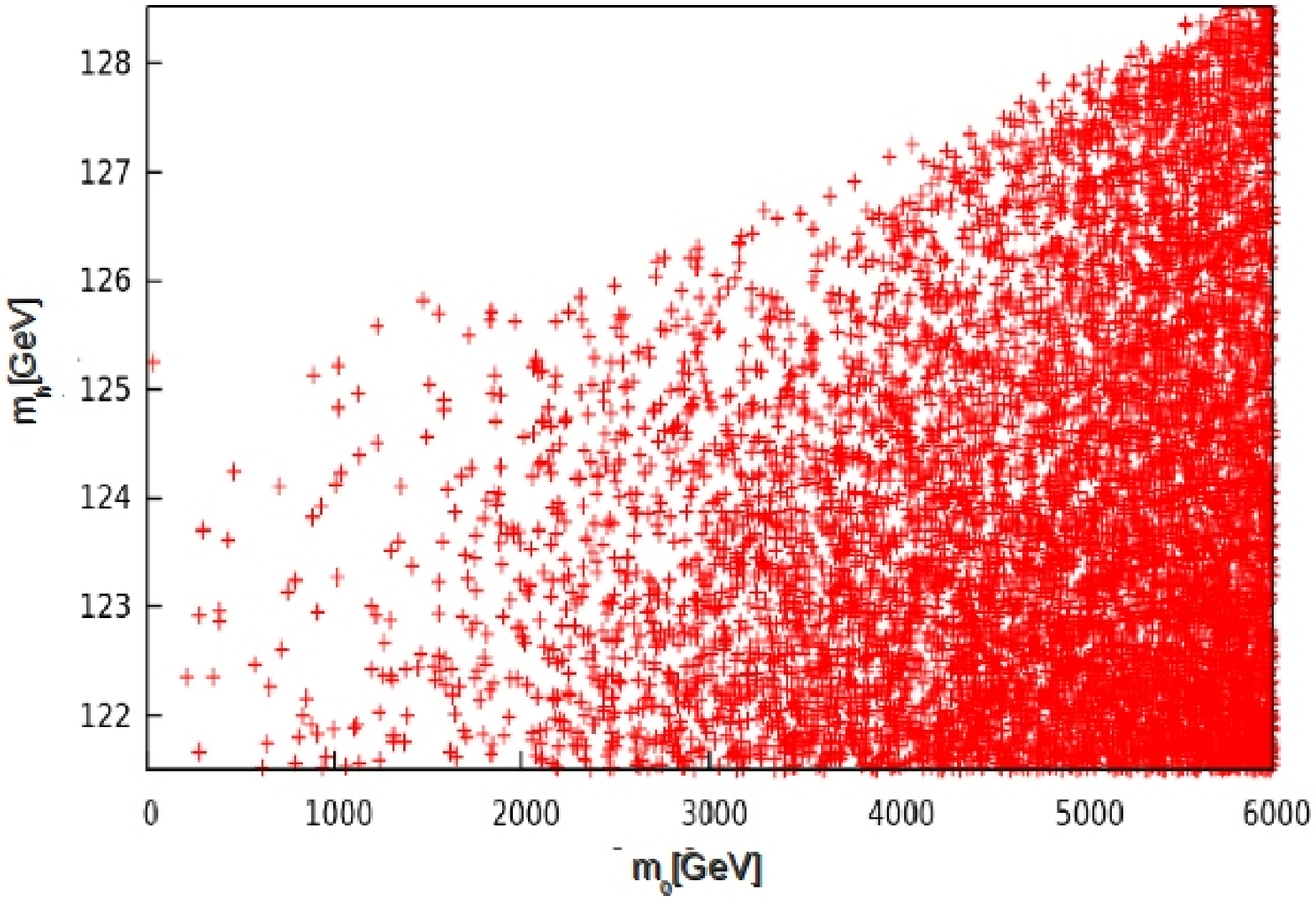}}\end{subfigure} 
\begin{subfigure}{\includegraphics[height=6cm,width=7.9cm]{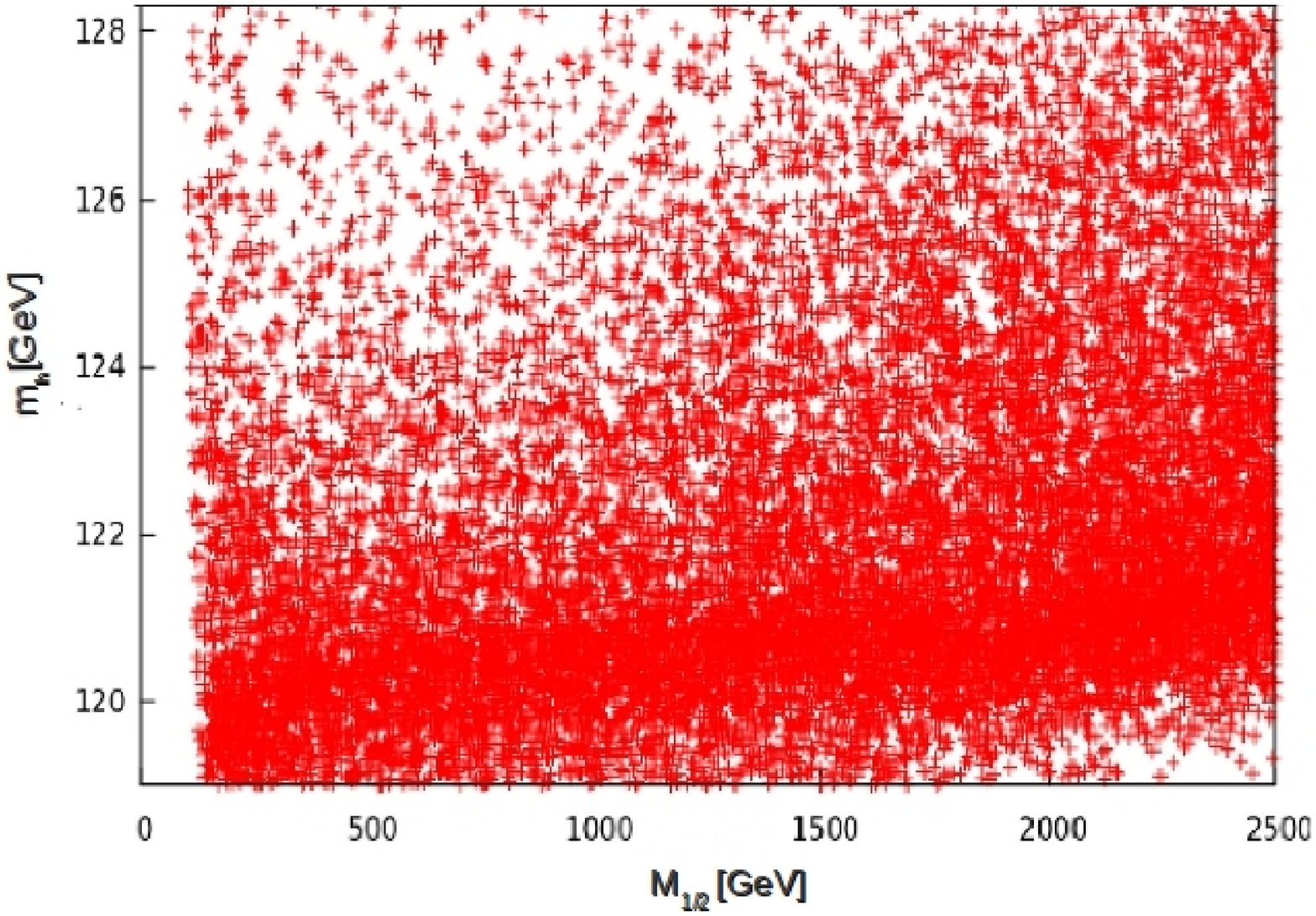}}\end{subfigure}\\
\centering{
{\includegraphics[height=6cm,width=7.9cm]{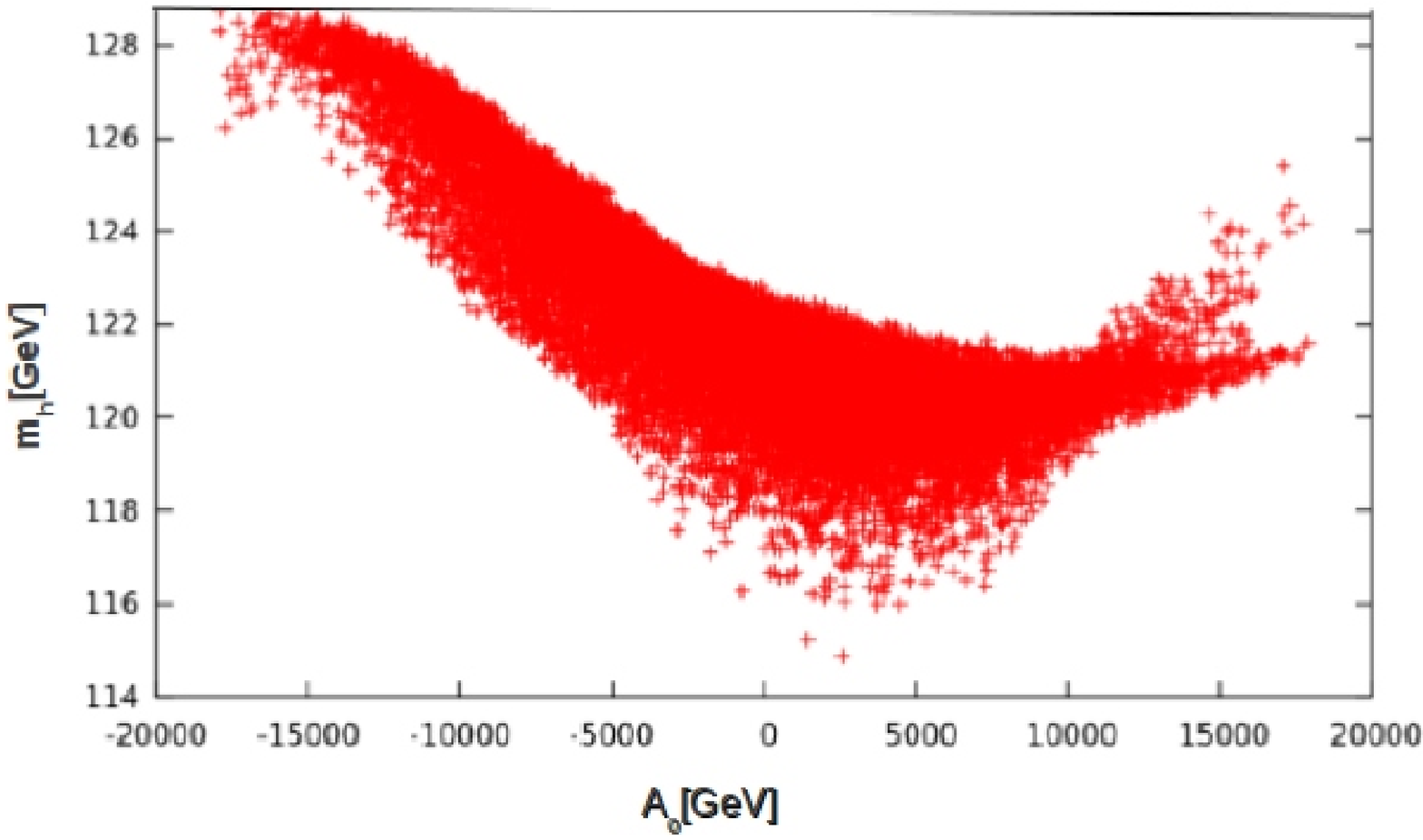}} \\}
\caption{The results of our calculations are presented for NUHM case. In \textbf{a}, different horizontal lines represent the present (MEG 2013) and future MEG bounds for BR($ \mu $ $ \rightarrow $ e + $ \gamma $). \textbf{b-e} shows the allowed space for different parameters, that is allowed by MEG 2013 bound.}
%\label{some example}
\end{figure}

\par
From Fig. 3a, we find that log[BR($ \mu \rightarrow e+\gamma$)] increases with increase in scalar masses (in contrast to mSUGRA and NUHM). This could be due to some strong cancelations occurring because of the particular ratios of gaugino masses in NUGM model as discussed earlier. In Fig. 3b, the SUSY parameter space  $m_{0} - M_{3}$ as allowed by the MEG 2013 bound on BR($\mu\rightarrow e \gamma$) is presented. We find that $M_{3} \geq$ 2 TeV  is allowed for almost the whole range of $ m_{0} $; while for low $ M_{3} \leq $ 2 TeV, smaller values of $ m_{0} $ are favored. The region below the curve line is excluded by SUSY.
\par 
In Fig. 4a, c, e we present the constraints from  $BR( \mu \rightarrow e+\gamma)$ on NUGM parameter space for tan$ \beta $ = 10. As can be seen, a large part of the paramater space survives for tan$ \beta $ = 10 in NUGM, as compared to NUHM and mSUGRA. From Fig. 4b we find that for Higgs mass $m_{h}$ around 125.9 GeV, the whole parameter space $m_{0} \geq $ 1.5 TeV is allowed. Squark masses $m_{0} \geq $ 1.5 TeV corresponding to 126 GeV Higgs are mostly favored, which would be accessible at the next run of LHC (satisfying the current MEG limit $BR( \mu \rightarrow e+\gamma)\leq 5.7\times10^{-13}$). From Fig. 4d, we see that for a Higgs mass around 126 GeV, $ M_{1} $ lies between $ -2.8 \hspace{.1cm}\text{TeV} \leq M_{1}\leq -1 \hspace{.1cm} \text{TeV}$. The constraints imposed on the soft SUSY breaking parameters, in NUGM space, are found to be less severe compared to NUHM and mSUGRA. The plot of $ A_{0}[GeV] $ Vs $ m_{h} [GeV]$ is shown in Fig. 4f. The patches in the plot are due to cancelation in the entries of the left-handed slepton mass matrices $ \delta^{LL}_{i \neq j} $ between the soft universal mass terms.
\begin{figure}
\begin{subfigure}{\includegraphics[height=6cm,width=7.95cm]{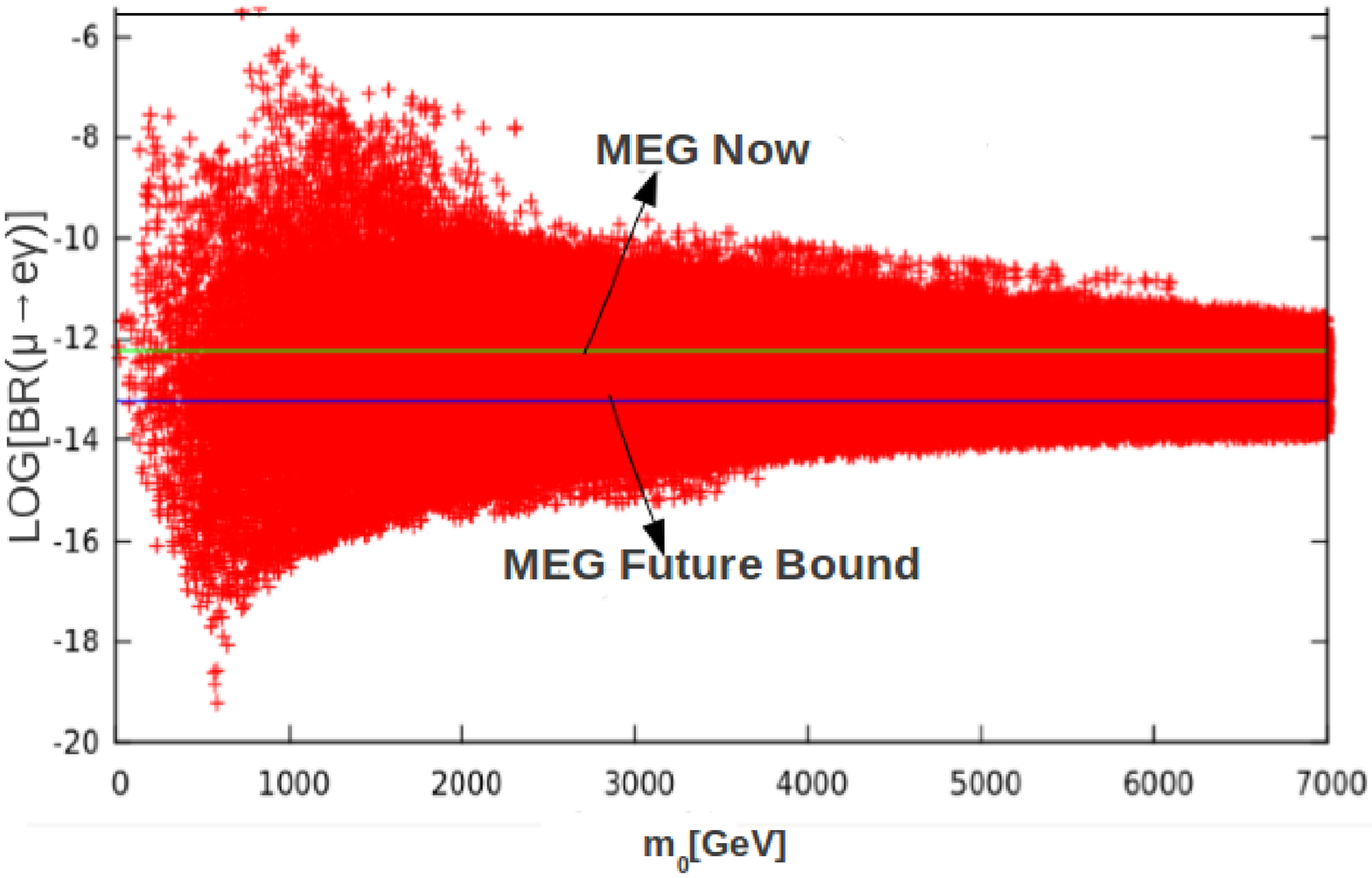}}\end{subfigure} 
\begin{subfigure}{\includegraphics[height=6cm,width=7.95cm]{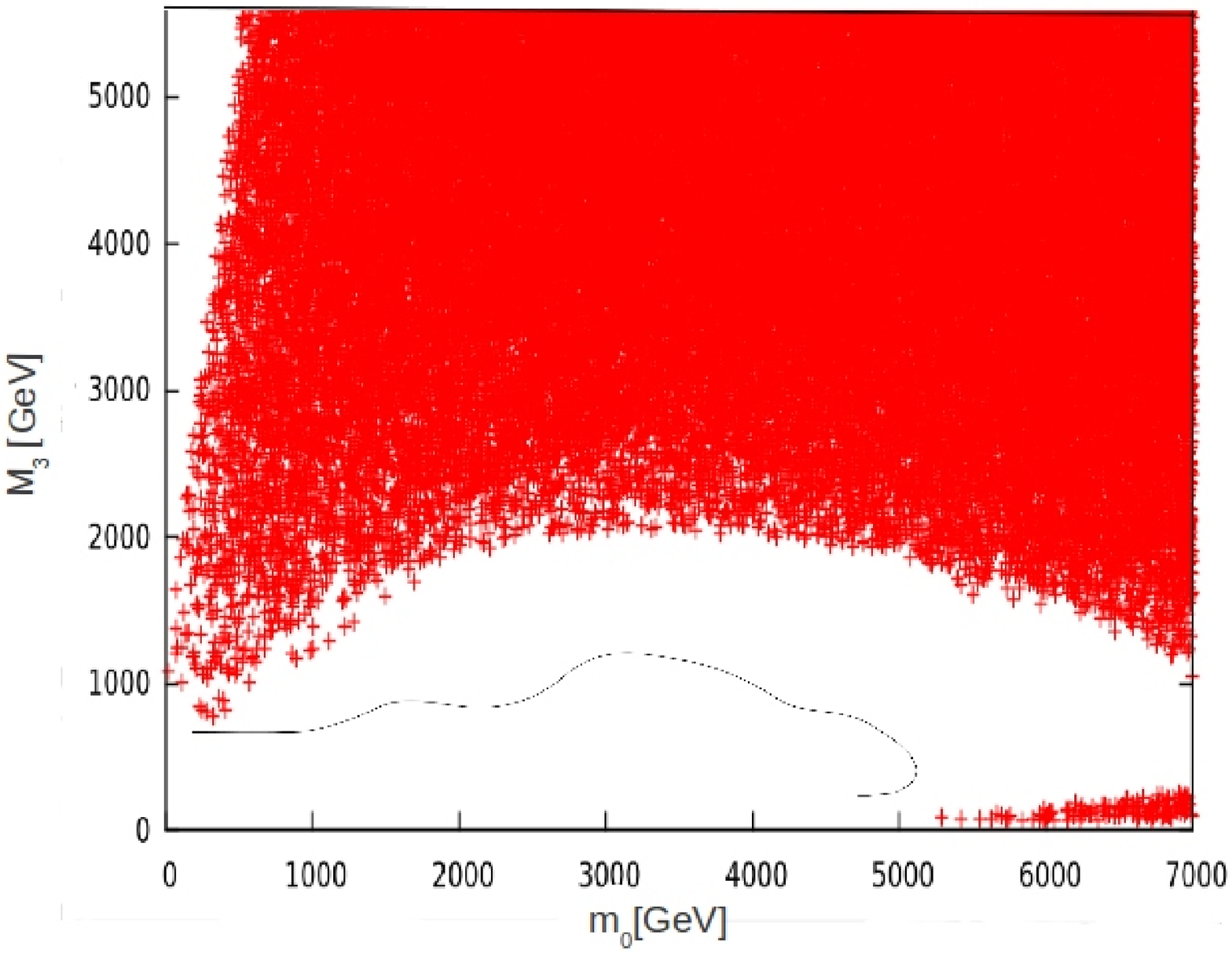}}\end{subfigure}
\caption{\textbf{a } we show the plot $m_{0}$ [GeV] vs. log [BR($ \mu \rightarrow e+
\gamma$)], \textbf{b} represents parameter space of $m_{0}$ and $M_{3}$, for NUGM model.
 Different horizontal lines represent present and future bounds on BR($\mu\rightarrow e \gamma$).}
%\label{some example}
\end{figure}

\subsection{\textbf{Non Universal Scalar Mass Models (NUSM)}}
\bigskip
The parameters of NUSM model are given by \cite{Chatto},
\par 
$\hspace{.7cm}$ $\text{tan}\beta$, $M_{1/2}$, $A_{0}$, $\text{sgn}(\mu)$, \text{and} $m_{0}$. 
\newline
The parameters play exactly the same role as those in mSUGRA, except for a significant difference in the scalar sector. The masses of the first two generations of scalars
(squarks and sleptons) and the third generations of sleptons are designated as $ m_{0} $ at the GUT scale. Here $ m_{0} $ is allowed to span up to a very large value of up to tens of TeVs. However, the Higgs scalars and the third family of squarks are assumed to have vanishing mass values at $ M_{GUT} $. In this analysis the mass parameters for the third generation of squarks and Higgs scalars are set to zero. We limit ourselves to a vanishing $ A_{0} $ in our analysis \cite{S.Bhatt}. We present our results obtained with the non-universal scalar masses at $ M_{GUT} $ in Fig. 5. In Fig. 5a we show the SUSY parameter space as allowed by the present and future MEG bounds on $BR(\mu \rightarrow e \gamma)$. For the present MEG bound on log $BR(\mu \rightarrow e \gamma)$, the allowed region of $ M_{1/2} $ parameter space becomes constrained with a lower limit of 400 GeV. Figure 5b in the right panel shows $ m_{0} $ [GeV] Vs $ M_{1/2} $ [GeV] as allowed by the MEG 2013 bound on BR$(\mu \rightarrow e \gamma)$ (The values of $m_{0}$ (GeV) along the x-axis in Fig. 5b, c are multiplied by $ 10^{4} $ ). We find that the $ M_{1/2} \geq $ 1.2 TeV and $ m_{0} \geq $ 2.0 TeV region is allowed. From Fig. 5c we find that for a Higgs boson mass around 125 GeV, $ m_{0} \geq $ 2 TeV is favored. From Fig. 5d we see that for a Higgs mass around 125 GeV, $ M_{1/2} $ lies between 4 TeV $ \leq M_{1/2} \leq $ 6 TeV. The two processes $\mu \rightarrow e \gamma$ and $\mu \rightarrow 3e$ are correlated as can be seen from Fig. 5e. 
\par 
The variation of BR($\mu\rightarrow e\gamma$) with $ \frac{m_{1}-m_{2}}{m_{1}+m_{2}}$  is shown in Fig. 6, where $m_{1}$ and $ m_{2} $ are the masses of the first and second generation sfermions, respectively. The range of $\frac{m_{1}-m_{2}}{m_{1}+m_{2}}$ is taken to be from -0.1 to 0.1. The value of log[$BR(\mu \rightarrow e \gamma)$] varies from −14.8 to −14.9 for the given interval of $\frac{m_{1}-m_{2}}{m_{1}+m_{2}}$, and this change is quite insignificant. We find that the low energy flavor phenomenology is not much affected by these completelynon-universal SO(10) symmetric mass terms at the GUT scale.
\par 
We find that in CMSSM/mSUGRA like models, the present experimental limit on $BR(\mu \rightarrow e \gamma)$ disfavors the soft SUSY breaking parameters $ m_{0} \leq $ 6 TeV and $ m_{0} \leq $ 6 TeV if the Dirac neutrino Yukawas are used from the $\mu$-$\tau$
symmetric SUSY SO(10) theory \cite{Joshipura}. The LFV constraint
on the SUSY spectrum is relaxed if the NUHM model is considered and we find that an interesting cancelation in the magnitude of charged LFVs arises if the universality condition is relaxed for the soft mass of up type Higgs $ m^{2}_{H_{u}} $ . As a result of this, as compared to mSUGRA, a relatively soft parameter space is allowed in NUHM, by the BR($\mu \rightarrow e \gamma$) bounds. In mSUGRA if the seesaw scale is lower than the GUT scale, mixings take place among the sleptons of a different generation at the seesaw scale through (i) renormalization group evolution (RGE) effects and (ii) lepton flavor violating Yukawa couplings. As a result, the slepton mass matrices no longer remain diagonal at the seesaw scale. At the weak scale, the off-diagonal entries in the slepton mass matrices generate LFV decays. These effects have been studied in the
literature in all three variants of the seesaw mechanisms \cite{Rossi, F. Joaqium, FJ, F.Joaq, Arganda}.
\par 
\bigskip
In Tables III and IV we have summarized the comparison of our study with \cite{L.Cabbibi2012}. In Tables V and VI we have highlighted the comparative study of our analysis between NUGM and NUSM. The new results in NUGM which we find in our work are the following:
\bigskip
\newline
1. Lighter $m_{0}$ is also allowed as compared to mSUGRA.
\newline
2. A wider SUSY parameter space is allowed.
\newline
3. The $A_{0}$ range in this work is shown in the Table 5.
\newline
4. BR($\mu\rightarrow e \gamma$) increases with increase of masses.
\bigskip

\begin{table*}
\caption{Masses in this table are comparison between \cite{L.Cabbibi2012} and this work for NUHM.}
\label{sphericcase}
\begin{tabular*}{\textwidth}{@{\extracolsep{\fill}}lrrrrl@{}}
\hline\noalign{\smallskip}
\hline
Range of parameters allowed by & Range of parameters allowed by
\\  for BR $\left( \mu \rightarrow e \gamma\right) < 5.7 \times 10^{-13}$  & BR $\left(\mu \rightarrow e\gamma\right) < 2.4\times 10^{-12}$  
\\  MEG 2013 (from this work) & MEG 2012 (L. Cabbibibi et al.\cite{L.Cabbibi2012}) \\
 \hline
 \textbf{1}.Figure 2a: & \textbf{1}.Only $M_{1/2}\geq$ 0.5 TeV 
\\almost whole $M_{1/2}$ space allowed \\ 
\textbf{2}.Figure 2b: (MEG 2013) & \textbf{2}.$m_{0}\geq$ 3 TeV for small  
\\$m_{0}\geq$ 1.5 TeV for &  $ M_{1/2} $, $M_{1/2} \geq$ 1 TeV   
\\ $M_{1/2} \geq$ 500 GeV &  for small $ m_{0}$
\\wider space is allowed in this work. \\
\textbf{3}.Figure 2d:& \textbf{3}.$m_{0}\geq$ 3.2 TeV for
\\$m_{0}\geq$ 2.3 TeV for & $ m_{h} = 125.9  $ GeV
\\$m_{h} = 125.9$ GeV\\ 
\textbf{4}.Figure 2e: & \textbf{4}.Almost same as in ours.
\\ -13 TeV $ < A_{0} < $ -7 TeV 
\\ for $ m_{h} = 125.9$ GeV\\
\hline
\end{tabular*}
\end{table*}
\begin{table*}
\caption{Masses in this table are comparison between NUGM and NUSM of this work.}
\label{sphericcase}
\begin{tabular*}{\textwidth}{@{\extracolsep{\fill}}lrrrrl@{}}
\hline\noalign{\smallskip}
\hline
Range of parameters allowed by & Range of parameters allowed by
\\ for BR $\left( \mu \rightarrow e \gamma\right) < 5.7 \times 10^{-13}$ & for BR $\left(\mu \rightarrow e\gamma\right) < 5.7\times 10^{-13}$\\ in NUGM  & in NUSM \\
 \hline
 \textbf{1}.Figure 4e: & \textbf{1}.Figure 5a:\hspace{.1cm}constrained region of $ M_{1/2} $
\\almost whole $M_{3}$ space is allowed, & parameter space is allowed,\\
100 GeV $ \leq M_{3}\leq$ 5.8 TeV. & $ M_{1/2}\geq $ 400 GeV.\\ 
\textbf{2}.Figure 3b: & \textbf{2}.Figure 5b:\hspace{.1cm}$M_{1/2}\geq$ 1.2 TeV for 
\\$M_{3}\geq$ 2 TeV for whole & whole $ m_{0}$, $m_{0} \geq$ 2 TeV  
\\ $m_{0} \leq$ 1 TeV, $M_{3} \leq$ 2 TeV, & for whole $ M_{1/2} $.  
\\for small $ m_{0} \leq $ 1 TeV.  \\
\textbf{3}.Figure 4b:& \textbf{3}.Figure 5c:\hspace{.1cm}$m_{0}\geq$ 2 TeV
\\$m_{0}\geq$ 850 GeV for &  for $ m_{h} = $ 125 GeV.
\\$m_{h} = $ 125 GeV. \\ 
\textbf{4}.Figure 4d: & \textbf{4}.Figure 5d:\hspace{.1cm}4 TeV $\leq M_{1/2}\leq$ 6 TeV 
\\ $|M_{1}|\geq$ 900 GeV &  for $ m_{h} =$ 125 GeV. 
\\ for $ m_{h} = $125 GeV.\\
\hline
\end{tabular*}
\end{table*}
\begin{table*}
\caption{Masses in this table are comparison between \cite{L.Cabbibi2012} and this work for mSUGRA.}
\label{}
\begin{tabular*}{\textwidth}{@{\extracolsep{\fill}}lrrrrl@{}}
\hline\noalign{\smallskip}
\hline
Range of parameters allowed by & Range of parameters allowed by
\\  for BR $\left( \mu \rightarrow e \gamma\right) < 5.7 \times 10^{-13}$  & BR $\left(\mu \rightarrow e\gamma\right) < 2.4\times 10^{-12}$  
\\  MEG 2013 (from this work) & MEG 2012 (L. Cabbibibi et al.\cite{L.Cabbibi2012}) \\
 \hline
\textbf{1}.Figure 1a:& \textbf{1}.For  MEG 2011, $M_{1/2}\geq$ 0.5 TeV 
\\$M_{1/2}\geq$ 1 TeV by MEG 2013 & and  $M_{1/2} \geq$ 1.5 TeV  
 \\ and very heavy $M_{1/2} \geq$ 3.5 TeV & for BR$\left( \mu \rightarrow e \gamma\right)<10^{-13}$ 
  \\ by future MEG bound \cite{Adam}\\ 
\textbf{2}.Figure 1b (MEG 2013): & \textbf{2}. For MEG 2011 of BR $\left(\mu \rightarrow e\gamma\right)$,
\\ $m_{0}\geq$ 6 TeV for small $M_{1/2}$ & $m_{0}\geq$ 4 TeV for small $ M_{1/2}, $ 
\\ $M_{1/2} \geq$ 2 TeV for small $ m_{0}$ & $M_{1/2} \geq$ 2 TeV for small $ m_{0} $
\\$ M_{0} $ shifts to slightly heavier side \\
\textbf{3}.Figure 1c: & \textbf{3}.$m_{0}\geq$ 4 TeV, $ m_{h} = 125.9\hspace{.1cm} \text{GeV} $
\\ $m_{0}\geq$ 3 TeV 
\\ for $m_{h} = 125.9 $ GeV \\
\textbf{4}.Figure 1f: & \textbf{4}. -11 TeV $ < A_{0} < $ -6 TeV for  
\\ -12 TeV $ < A_{0} < $ -6 TeV &  $m_{h} = 125.9\hspace{.1cm}\text{ GeV}$
\\for $m_{h} = 125.9\hspace{.1cm} \text{GeV}$\\
\hline
\end{tabular*}
\end{table*}

\begin{table*}
\caption{Comparison of $A_{0}$ between mSUGRA, NUHM and NUGM (this work)}
\label{tab:5}       % Give a unique label
% For LaTeX tables use
\begin{tabular*}{\textwidth}{@{\extracolsep{\fill}}lrrrrl@{}}
\hline\noalign{\smallskip}
 \ $ A_{0}$ (mSUGRA)& $A_{0}$ (NUHM) & $A_{0}$ (NUGM)
 \\ (TeV) &  (TeV) &  (TeV) \\
\noalign{\smallskip}\hline\noalign{\smallskip}
$-12 < A_{0} < -6$ & $-13 < A_{0} < -7$ & $-15 < A_{0}<-10$  \\
 \noalign{\smallskip}\hline
\end{tabular*}
\end{table*} 
\begin{figure}[htbp]
\begin{subfigure}[]{\includegraphics[height=6cm,width=7.9cm]{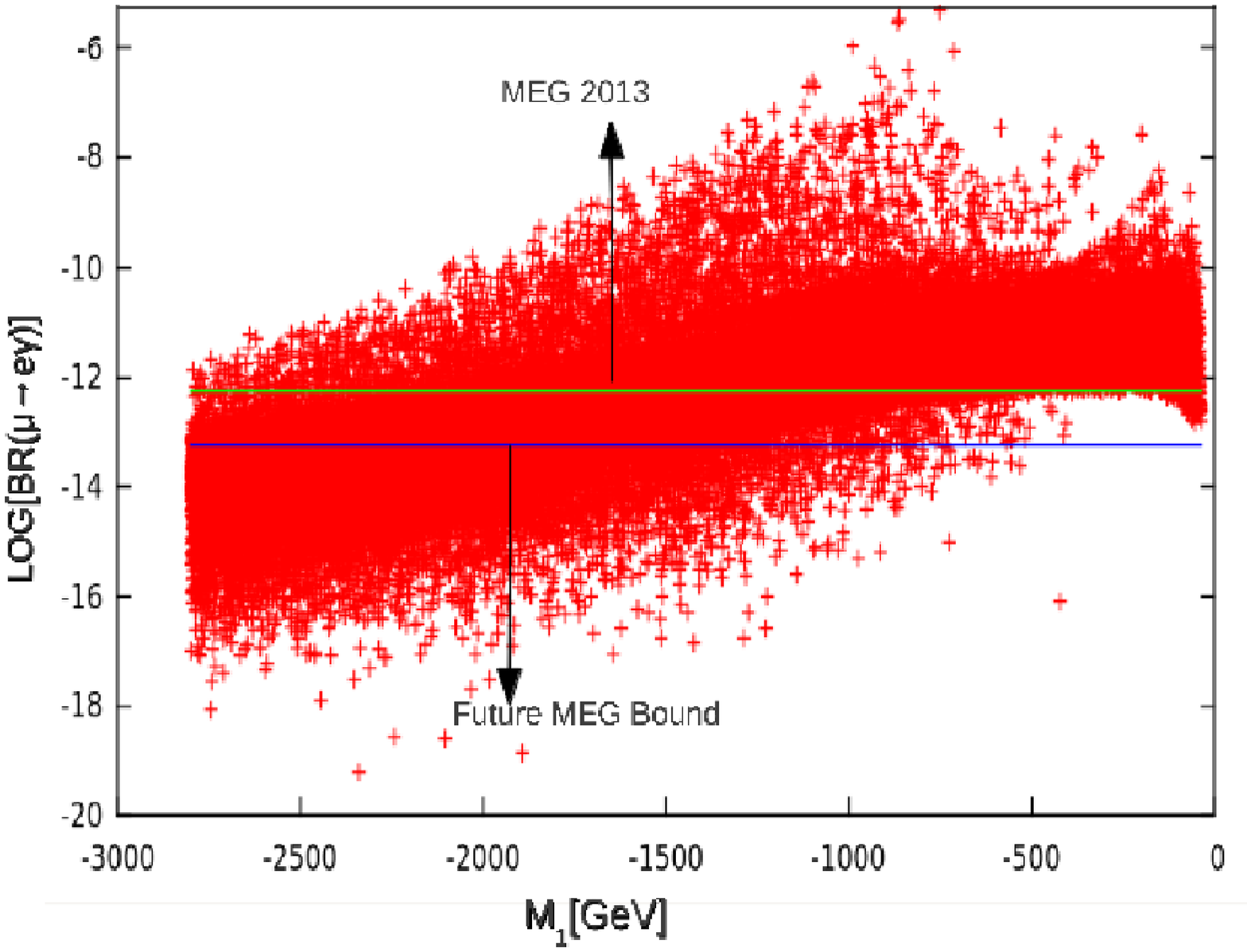}}\end{subfigure} 
\begin{subfigure}[]{\includegraphics[height=6cm,width=7.9cm]{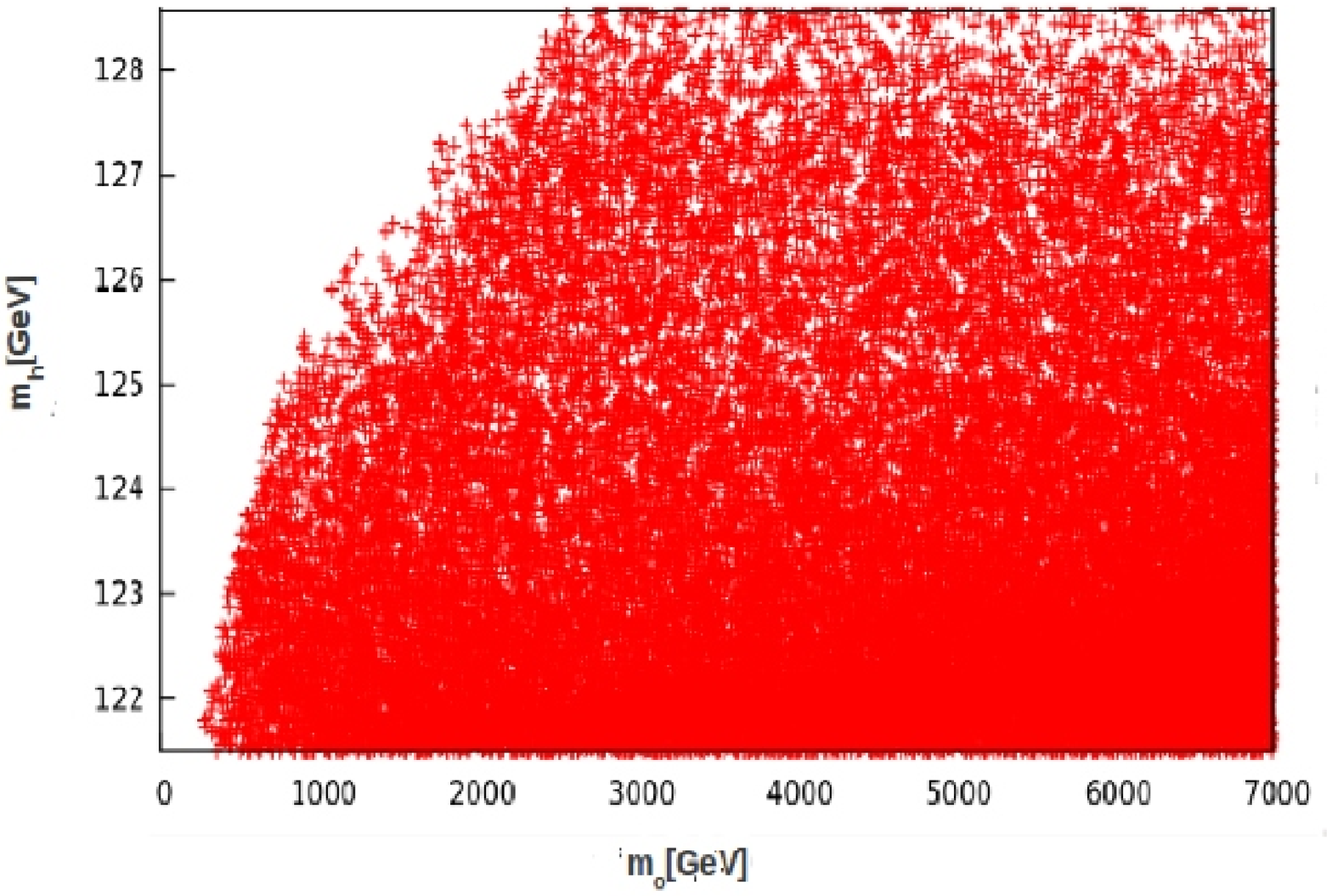}}\end{subfigure}\\
\begin{subfigure}[]{\includegraphics[height=6cm,width=7.9cm]{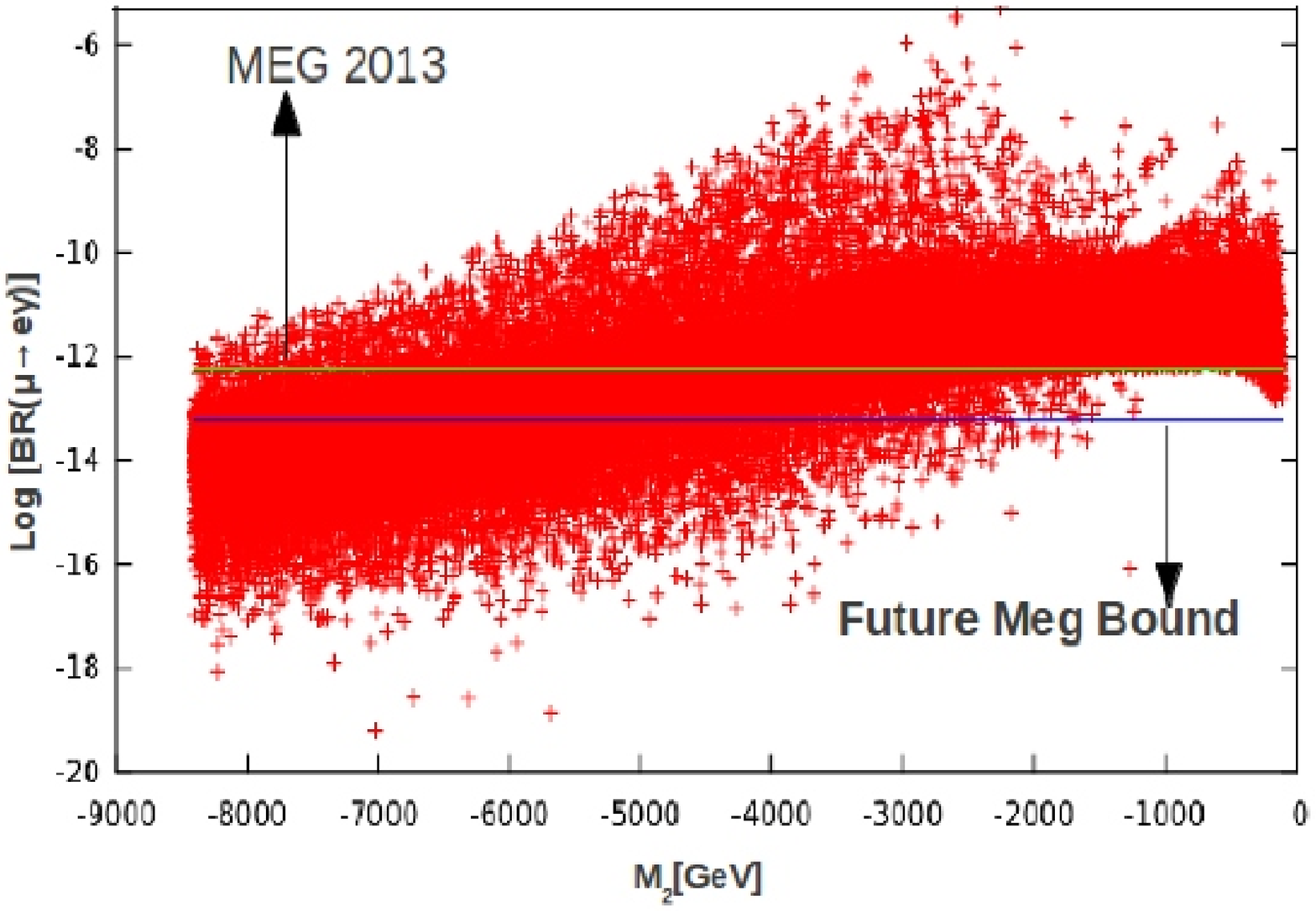}}\end{subfigure} 
\begin{subfigure}[]{\includegraphics[height=6cm,width=7.9cm]{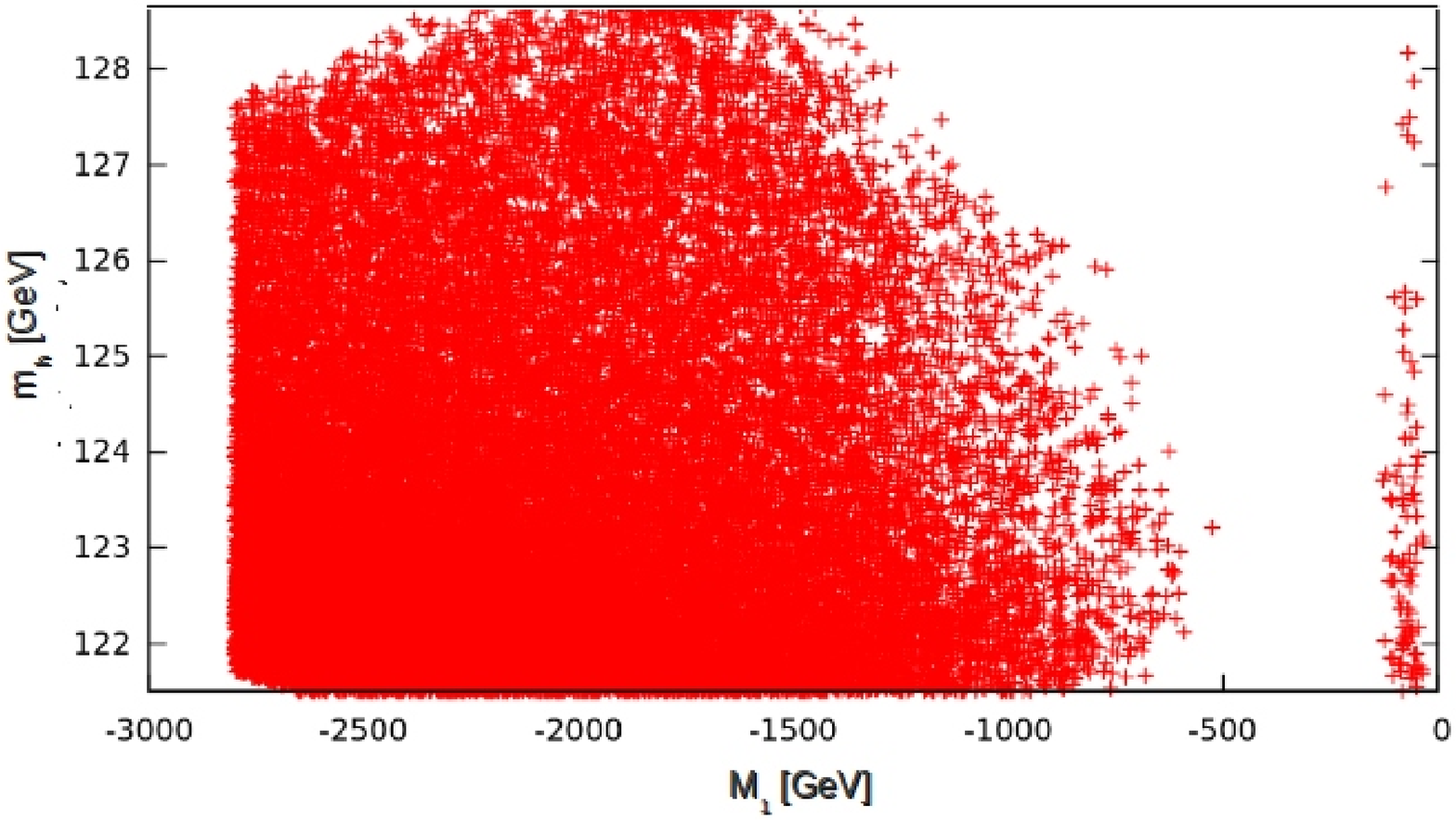}}\end{subfigure}\\
\begin{subfigure}[]{\includegraphics[height=6cm,width=7.9cm]{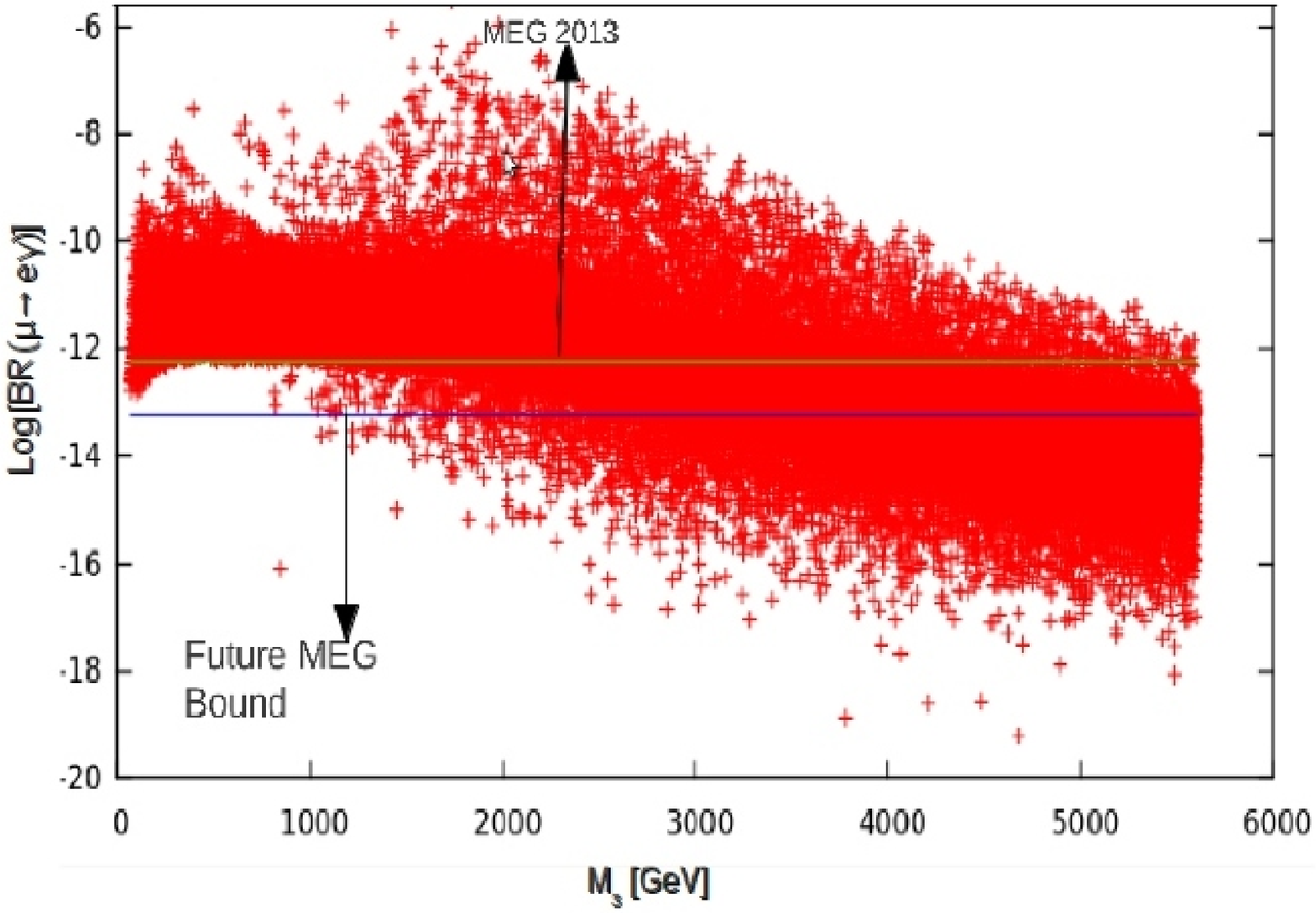}}\end{subfigure} 
\begin{subfigure}[]{\includegraphics[height=6cm,width=7.9cm]{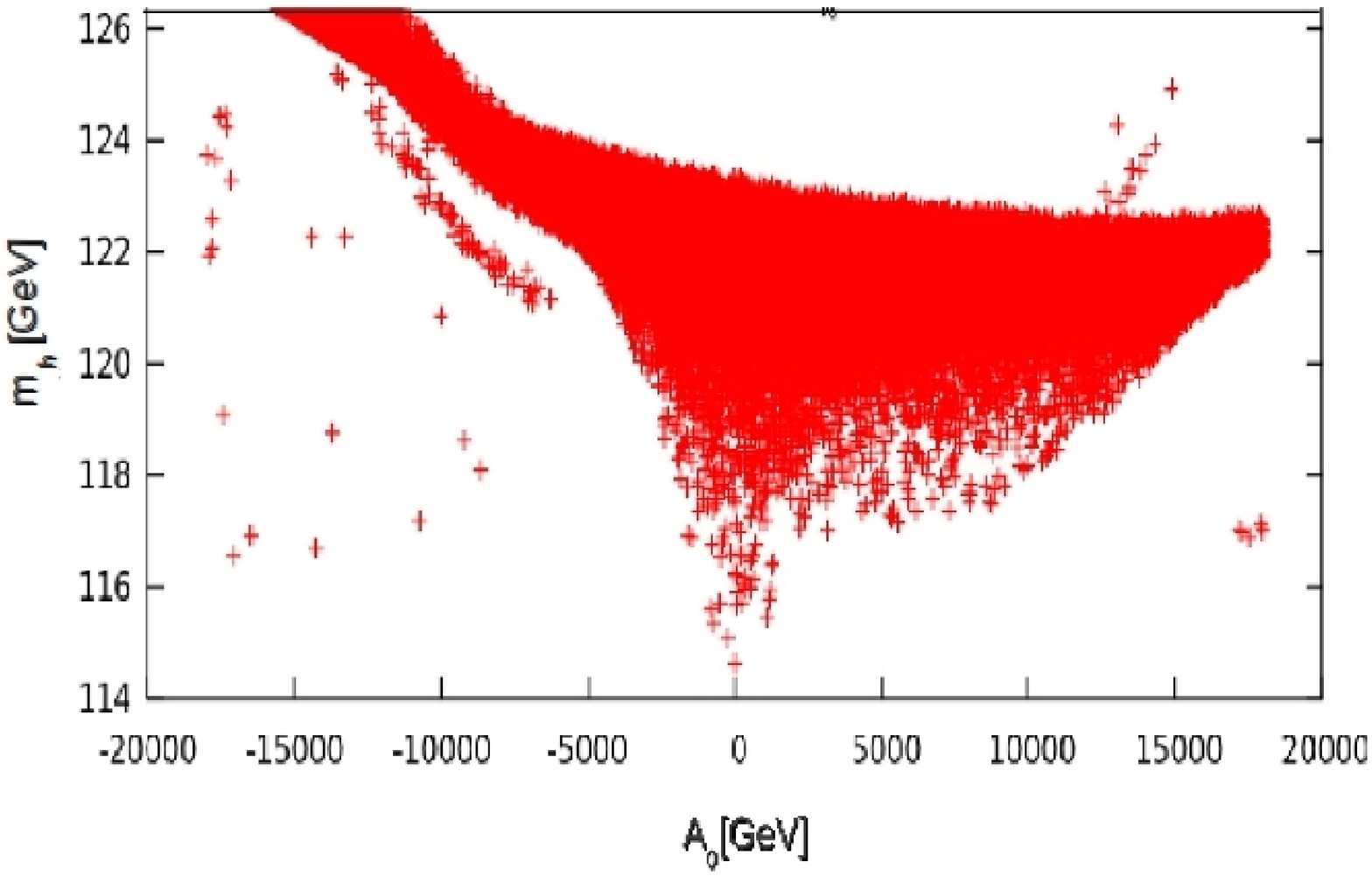}}\end{subfigure} \\
\caption{The results of our calculations are presented for NUGM case. In \textbf{a,c,e} different horizontal lines show the present (MEG 2013) and future MEG bounds for BR($\mu \rightarrow e + \gamma$). \textbf{b,d,f} The space for different parameters that is allowed by MEG 2013 bound.}
%\label{some example}
\end{figure}

\begin{figure}[htbp]
\begin{subfigure}[]{\includegraphics[height=6cm,width=7.9cm]{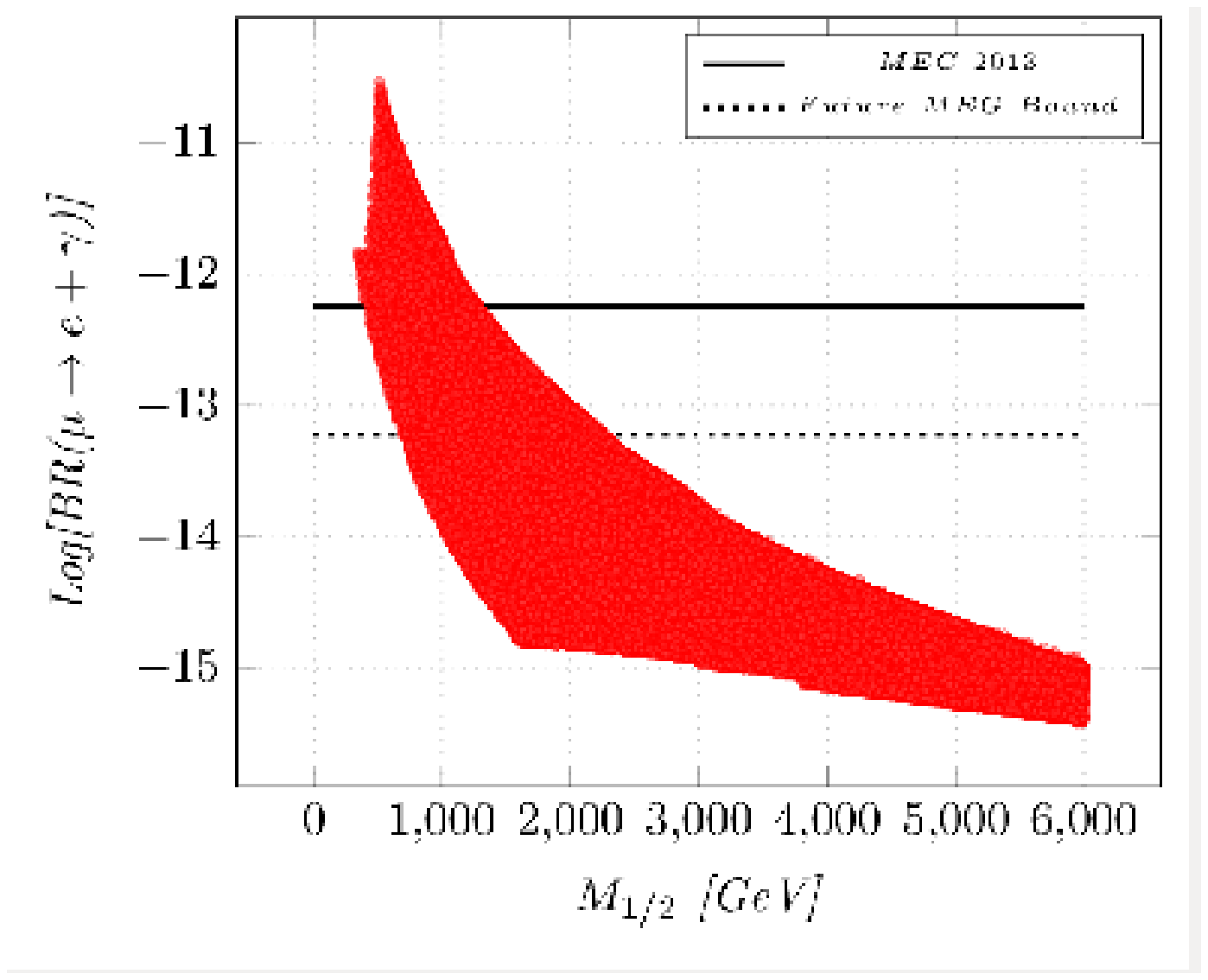}}\end{subfigure} 
\begin{subfigure}[]{\includegraphics[height=6cm,width=7.9cm]{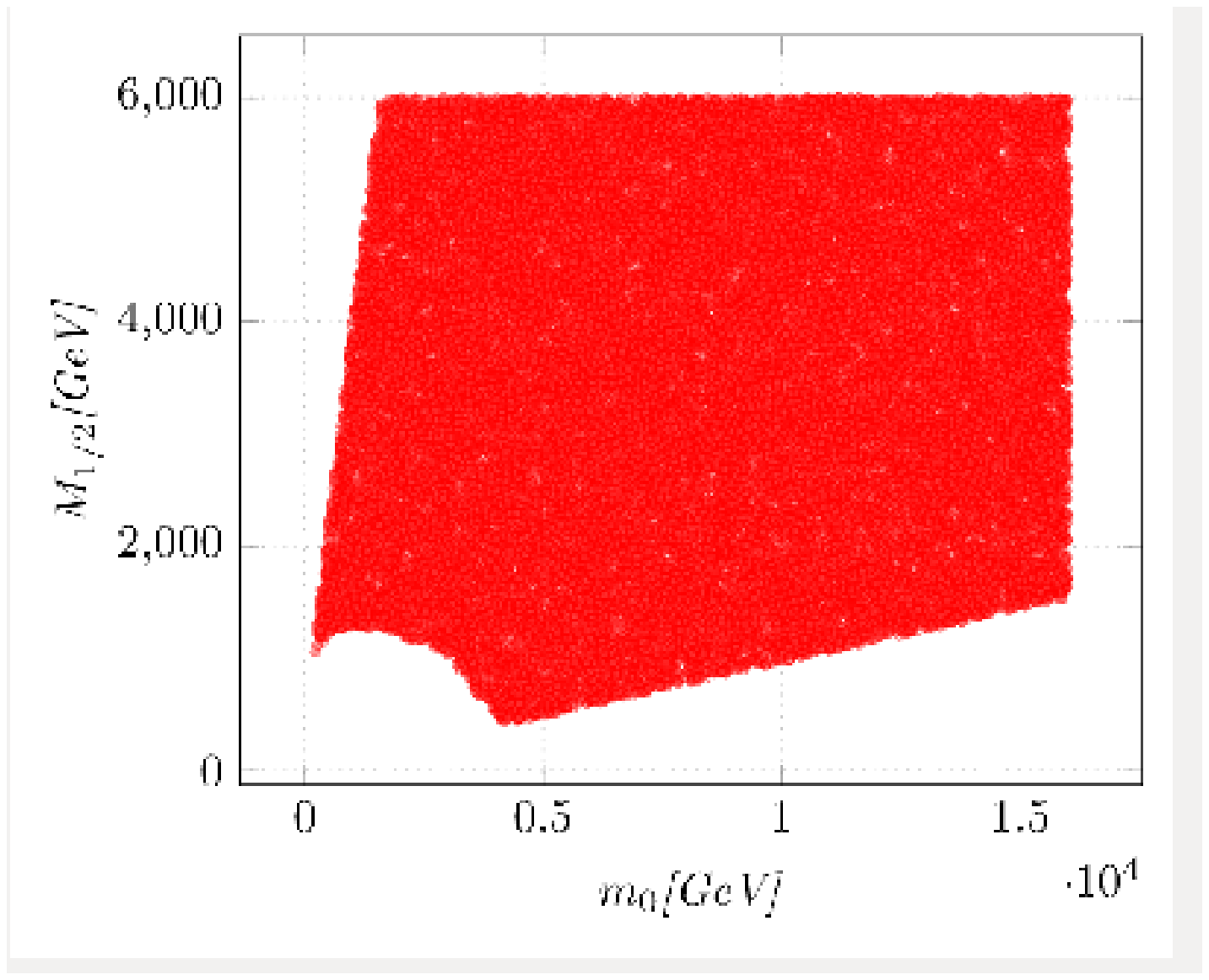}}\end{subfigure}\\
\begin{subfigure}[]{\includegraphics[height=6cm,width=7.9cm]{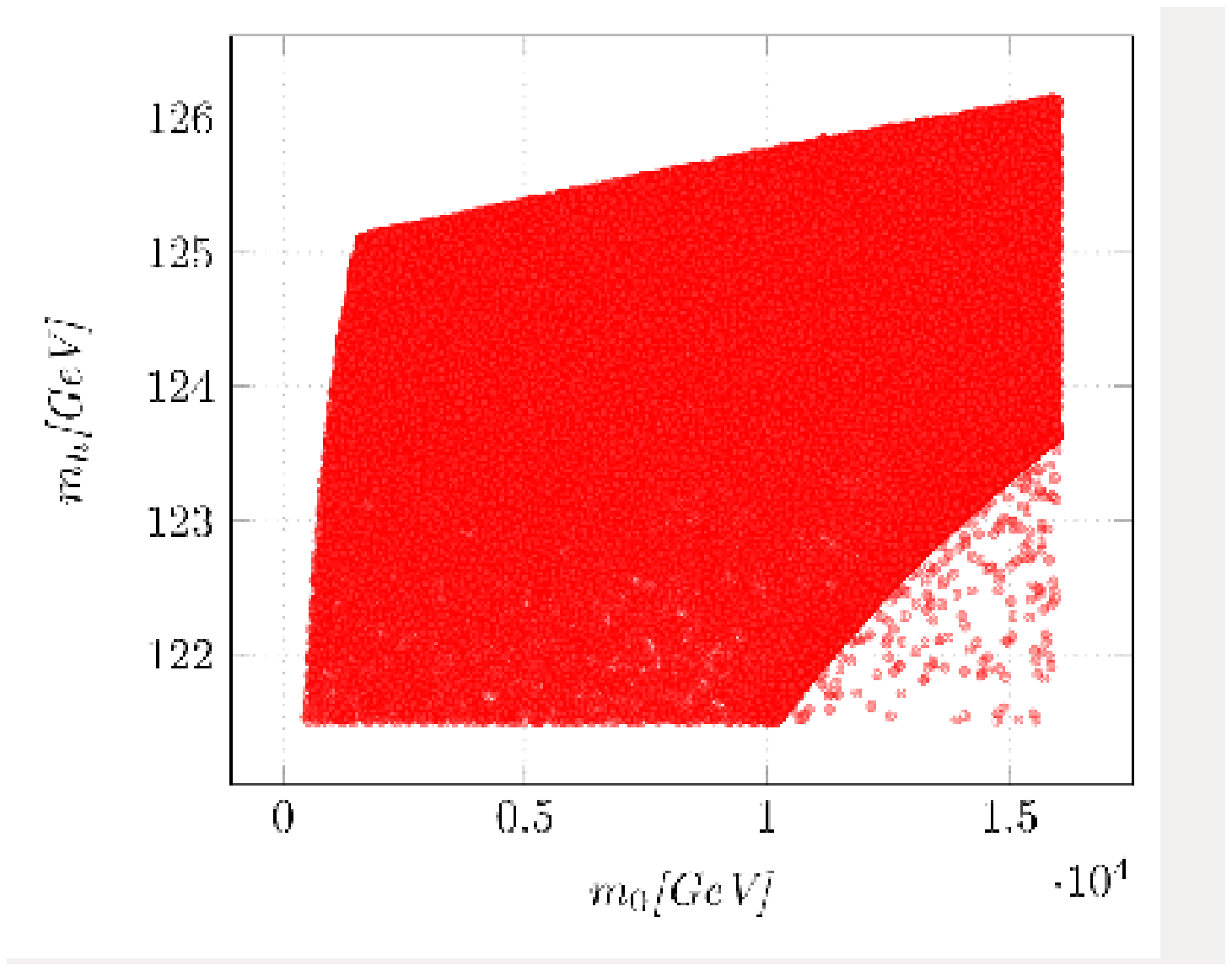}}\end{subfigure} 
\begin{subfigure}[]{\includegraphics[height=6cm,width=7.9cm]{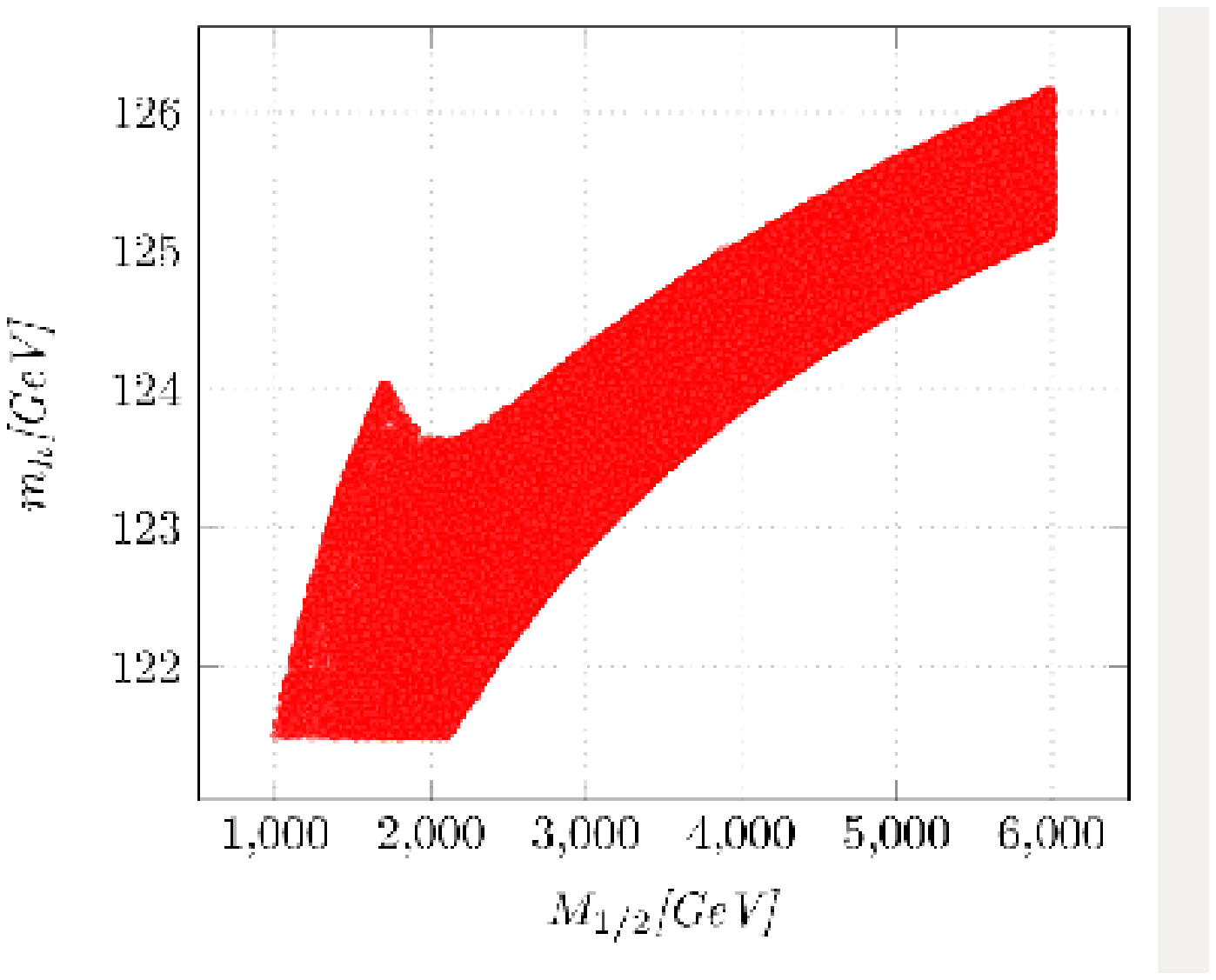}}\end{subfigure}\\
\centering{\includegraphics[height=6cm,width=7.9cm]{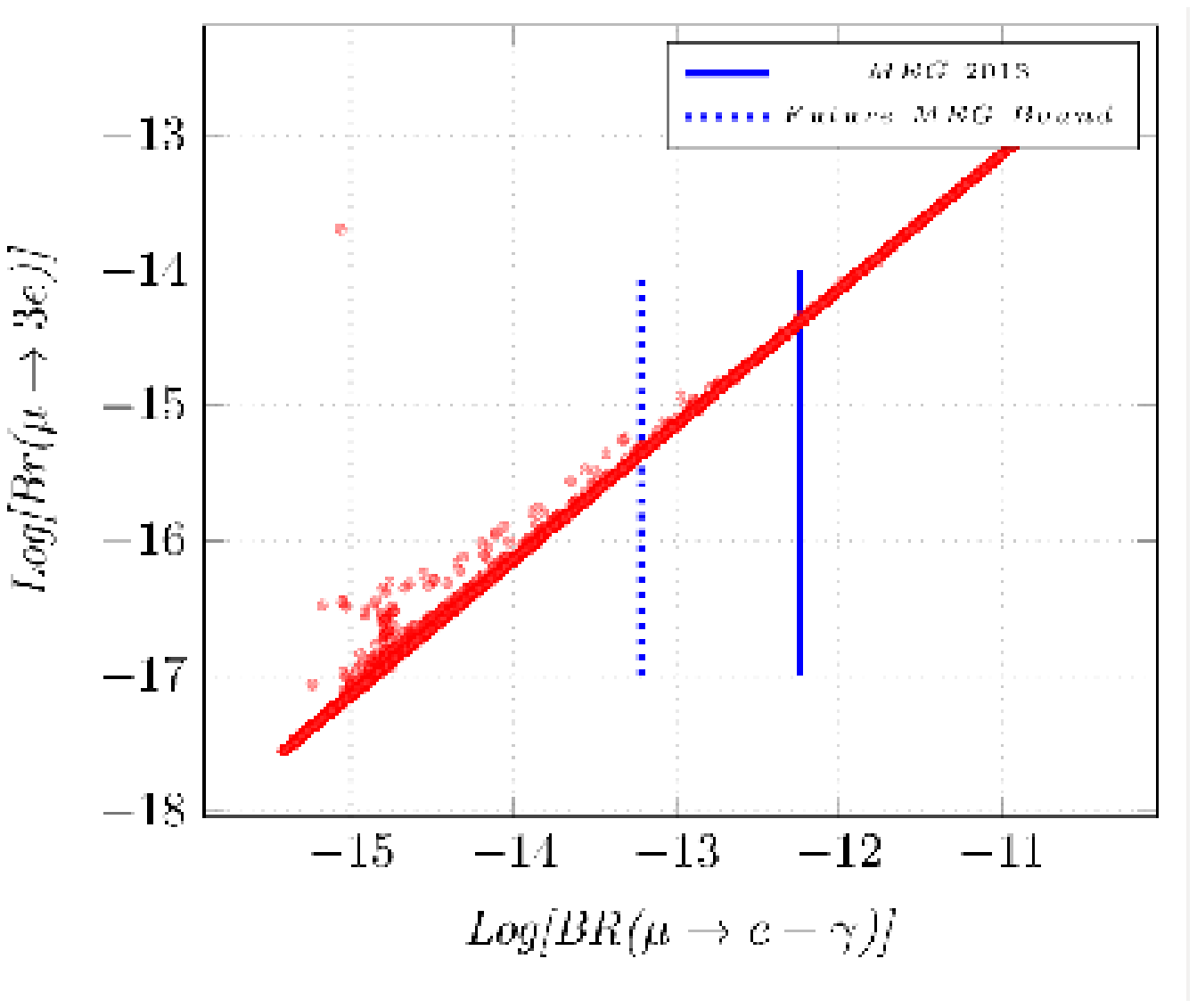}\\}
\caption{The results of our calculations are presented for NUGM case. In \textbf{a,c,e} different horizontal lines show the present (MEG 2013) and future MEG bounds for BR($\mu \rightarrow e + \gamma$). \textbf{b,d,f} The space for different parameters that is allowed by MEG 2013 bound.}
%\label{some example}
\end{figure}
 
\begin{figure}[htbp]
\centering{\includegraphics[height=6cm,width=7.9cm]{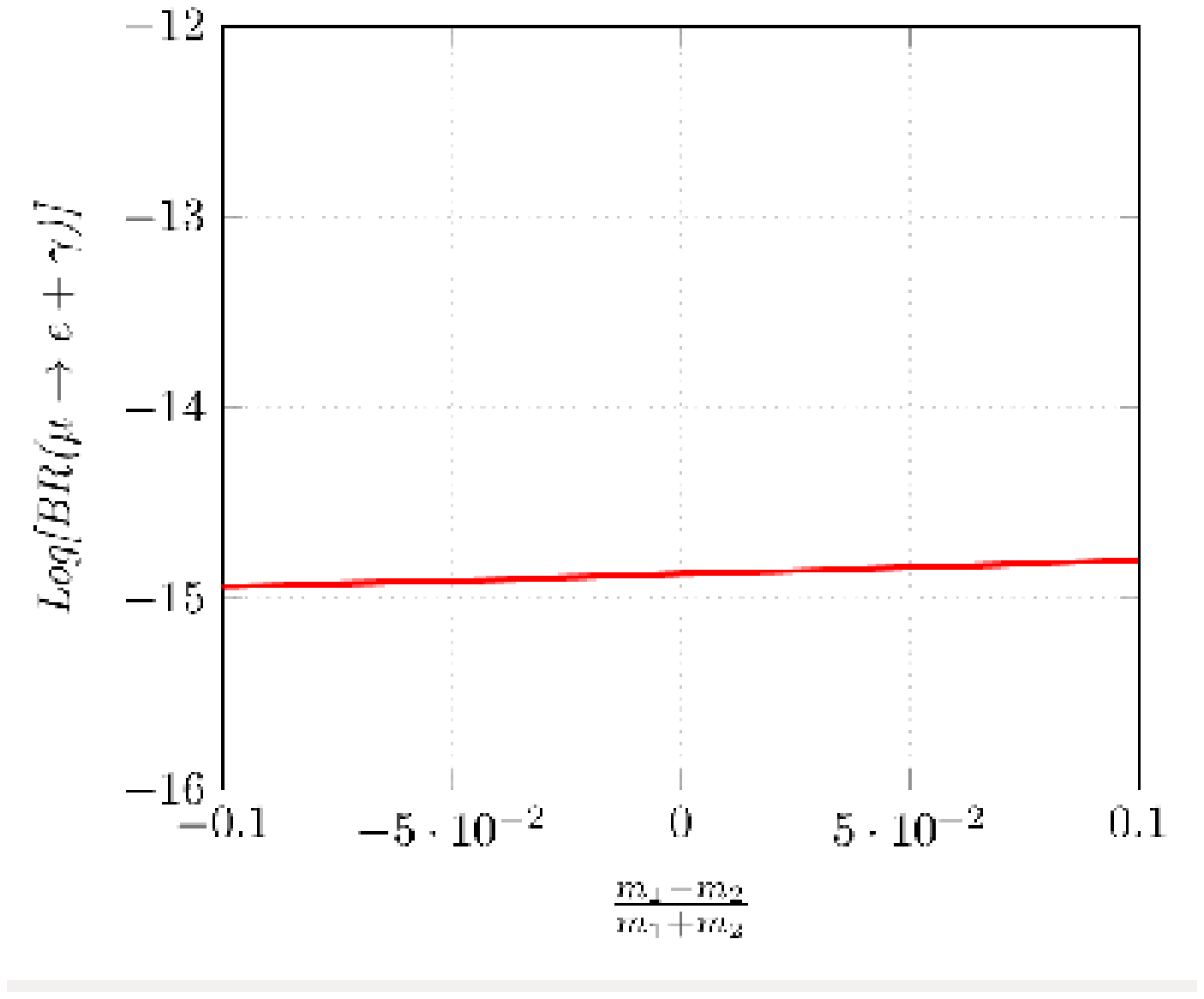}\\}
\caption{Variation of log[BR($\mu \rightarrow e + \gamma$)] as a function of $\frac{m_{1}-m_{2}}{m_{1}+m_{2}}$ is shown. The interval of  $\frac{m_{1}-m_{2}}{m_{1}+m_{2}}$ is taken to be from -0.1 to 0.1.}
\end{figure}
\section{Conclusion}
\bigskip
To conclude, in this work we have studied the rare cLFV decay $ \mu\rightarrow e\gamma $ in $\mu-\tau$ symmetric SUSY SO(10) theories, using the type I seesaw mechanism, in mSUGRA, NUHM, NUGM and NUSM models. We have used the value of the Higgs mass as measured at LHC, the latest global data on the reactor mixing angle $ \theta_{13} $ for neutrinos, and the latest constraints on BR($ \mu  \rightarrow e  \gamma $) as projected by MEG \cite{Adam, Baldini}. We find that in mSUGRA a very heavy $ M_{1/2} $ region is allowed by the future MEG bound of BR($ \mu  \rightarrow e  \gamma $), though in the NUHM case a low $ M_{1/2} $ is also allowed. Hence we further studied the non-universal gaugino mass model (NUGM). In mSUGRA, the $ m_{0} $ values as allowed by MEG 2013 bound, shift toward a heavier spectrum, as compared to allowed $ m_{0} $
of \cite{L.Cabbibi2012}(which was allowed by a less stringent bound of MEG
2011). As compared to mSUGRA, in NUHM, a wider parameter range is allowed. For a Higgs mass central value 125.9 GeV, our analysis allows a slightly lower value of $ m_{0} $ than \cite{L.Cabbibi2012}, both in mSUGRA and NUHM (as can be seen from
Tables III and IV). We find that NUGM allows, in general, a wider parameter space, as compared to both mSUGRA and NUHM. Here BR($ \mu  \rightarrow e  \gamma $) is found to increase with increase in $ m_{0} $ , which could be due to the particular ratios of gaugino masses. In NUGM, we find that the allowed values of $ \vert A_{0}\vert $ are shifted towards the heavier side (compared to mSUGRA and NUHM). In NUSM, the allowed $ M_{1/2} $ parameter space at low energies becomes constrained as compared
to the other three models. For a Higgs boson mass around 125 GeV, $ M_{1/2} $ lies between 4 TeV $ \leq M_{1/2} \leq $ 6 TeV and $ m_{0} \geq $ 2 TeV is mostly favored. The branching ratio of $ \mu  \rightarrow e  \gamma $ does not change significantly with variation of first and second generation sfermion masses at the GUT scale, in the
completely non-universal NUSM model.
\par
The results presented in this work can influence the experimental signatures for the production of SUSY particles and can motivate a special detector set up to guarantee that
the largest possible class of supersymmetric models lead to observable signatures at the present and future run of LHC. Hence any observation of heavy particles at the next run of LHC could help us understand and discriminate among these models, in reference to constraints put by cLFV decays. This in turn could contribute towards a better understanding of theories beyond the standard model.

\section{Acknowledgements}
\bigskip
KB and GG would like to thank Katen Patel for discussions, and for bringing this problem to their attention. They also thank Sudhir Vempati for critically reading the manuscript and fruitful discussions and suggestions at IISc Bangalore. They sincerely thank
the referee for very constructive and helpful comments and suggestions. KB would like to thank Gauhati University, and ICTP, Italy, for providing support to visit ICTP, where part of this work was carried out. She also thanks John March Russel, for discussions on Maximally SUSY theories, at ICTP. GG acknowledges support from CHEP, IISc, Bangalore, to visit the Institute, where a major part of this work has been done. She would also like to thank UGC, India, for providing an RFSMS fellowship to her.

\end{document}